\documentclass[article]{revtex4}
\usepackage{graphicx}
\def\be{\begin{eqnarray} &&} 
\def\nonu{\nonumber \\ &&} 
\def\ee{\end{eqnarray}} 

\usepackage{color}

\usepackage{soul}
\newcommand{\bm}[1] {\mbox{\boldmath{$#1$}}}
\newcommand{\blf}[1]{\bf  {\tilde #1}}

\def\sumint{\int \! \!\ \! \! \! \! \!\ \! \! \!\! \!\sum}

\def\bq{\begin{eqnarray}}
\def\eq{\end{eqnarray}}

\begin{document}
\title{Light-Front {spin-dependent} Spectral Function and Nucleon Momentum Distributions
for a Three-Body System}
\author{Alessio Del Dotto$^{a,b}$, Emanuele Pace$^c$, Giovanni Salm\`e$^a$,
 Sergio Scopetta$^d$}
 
\affiliation{$^a$ Istituto  Nazionale di Fisica Nucleare, Sezione di Roma,  P.le A. Moro 2,
 I-00185 Rome, Italy,\\
 {$^b$ Thomas Jefferson National Accelerator Facility, Newport News, VA 23606, USA,\\}
{$^c$ Dipartimento di Fisica, Universit\`a degli Studi di Roma ``Tor Vergata''} and INFN, Sezione di  Roma 2, 
Via della Ricerca Scientifica 1, 00133 Rome, Italy, \\
%$^d$Istituto  Nazionale di Fisica Nucleare, Sezione di Roma, P.le A. Moro 2, I-00185 Rome, Italy,\\
{$^d$ Dipartimento di Fisica e Geologia, Universit\`a degli Studi di Perugia} and INFN, Sezione di Perugia, Via Alessandro Pascoli,
06123 Perugia, Italy.\\
}

\begin{abstract}
Poincar\'e covariant definitions for the {spin-dependent} spectral function and for the momentum {distributions} within the
light-front Hamiltonian dynamics are proposed for a three-fermion bound system, starting from the light-front 
wave function of the system.  The adopted approach is based on the Bakamjian-Thomas construction of the Poincar\'e 
 generators, that allows one to easily import
the familiar  and {wide} knowledge on  the  nuclear interaction into a
 light-front framework. The proposed  formalism can find 
 {useful applications} in  refined nuclear calculations, like the ones 
 needed for evaluating  the 
EMC effect or 
the semi-inclusive deep inelastic cross sections with polarized nuclear 
targets,
{since remarkably the light-front {unpolarized} momentum distribution  by definition 
 fulfills  both normalization and   momentum sum {rules}. It is also shown
%  how much 
{a  straightforward }
%is the 
generalization of the definition of the
 light-front spectral 
 function to an A-nucleon system}.
\end{abstract}

 \maketitle
 \section{Introduction}
\label{intro}
In {the analysis of} the next generation of high-energy 
{electron-nucleus} scattering experiments, planned
at the Jefferson Laboratory (JLab) upgraded at 12 GeV \cite{J12}, 
as well as at the 
future Electron-Ion Collider \cite{eic}), 
{refined description of nuclei}   will play a relevant role \cite{ds}, 
with a particular interest to the polarized
$^3$He target {at JLab12.}
High precision experiments, involving both protons and neutrons, are in 
fact necessary to clarify the flavour dependence of {(i)}
Parton Distribution Functions (PDFs), measured
in inclusive Deep Inelastic Scattering (DIS), 
and {(ii)} Transverse-Momentum Dependent Parton Distribution
(TMDs, see, e.g., Ref. \cite{bdr} for a general introduction),
accessed through Semi Inclusive DIS (SIDIS).
In the next few years, several experiments involving an $^3$He nuclear target 
will be performed at {JLab12}, with the aim at extracting information on the 
parton structure of the neutron.
New DIS measurements are planned \cite{emcnew,mar} and, in particular,
the three-dimensional neutron structure in momentum space, 
described in terms of quark TMDs, 
will be probed through SIDIS off polarized $^3$He, 
where a high-energy pion {(kaon)} is detected in coincidence with the scattered 
electron \cite{sid1,sid2}. 
%}

To be able to extract {{PDFs and TMDs in the neutron from DIS and SIDIS}} 
off $^3$He, accurate theoretical descriptions of the structure of $^3$He and 
of the scattering process are also needed.
{{Initial studies of DIS and SIDIS off $^3$He were performed in 
Refs. \cite{Pace} and
\cite{mio}, respectively}}, 
where the plane wave impulse approximation (PWIA) was adopted to describe the 
reaction mechanism, namely
the interaction in the final state (FSI) was considered only within the 
two-nucleon spectator pair which recoils. 
The $^3$He structure was treated non-relativistically, using the AV18 
NN interaction \cite{AV18}. 

In a recent paper \cite{Kaptari}, the spectator SIDIS process off polarized 
$^3$He, where a deuteron in the final state is detected, 
was studied taking into account for the first time the FSI between 
the hadronizing quark and the detected deuteron  
through a distorted spin dependent spectral function of $^3$He. The study of the standard SIDIS process {{off transversely
polarized}} $^3$He with a fast detected pion 
including the FSI is presented in Ref. \cite{tobe}, 
where the FSI between the observed pion and the remnant is again taken 
into account through a distorted spin-dependent spectral function {(preliminary results can be found in
 Ref. \cite{Dotto1})}. However, the description of the internal nuclear dynamics in  
\cite{Dotto1,tobe}, is still non-relativistic, or more appropriately non
 Poincar\'e covariant, while the high 
energies 
involved in the forthcoming {SIDIS experiments   \cite{sid1,sid2} should }require 
 at least a proper 
treatment of Poincar\'e covariance.

% In the last decades there has been a lot of efforts to get information on the  transverse 
% parton momentum distributions (TMDs)  \cite{boer,trento,mu-ta,Collins,sivers} through measurements of 
% semi-inclusive deep inelastic (SIDIS) electron scattering cross sections off nucleons 
% (see e.g.  \cite{06010,SIDIS}) or through  Drell-Yan experiments (see e.g.  \cite{COMPASS}).

In this paper, {the structure of a spin 1/2 three-nucleon system  will be
investigated within a relativistic, Poincar\'e covariant framework  (see Refs. \cite{Dotto,Pace1} 
{for early studies}). Indeed, our approach can be straightforwardly generalized to other spin 1/2 three-body
systems and even to complex nuclei}.
{To develop  a Poincar\'e covariant framework  that allows one to embed } the large amount of knowledge 
on the nuclear interaction obtained from the non relativistic description of nuclei, we adopt the Relativistic 
Hamiltonian Dynamics (RHD) \cite{Dirac} with a fixed number of on-mass-shell constituents in its light-front 
(LF) version \cite{KP,PG,coester,Lev1}. {Within the LF 
form of RHD, the Poincar\'e group 
  has a sub-group given by the  LF boosts}, which 
allows a kinematical separation of the intrinsic motion from the global one. {Such a
property  plays a very important role for the relativistic} description 
of DIS, SIDIS and deeply virtual Compton scattering (DVCS) processes, {where the final states {can} have a fast recoil}. Furthermore {
the {LF} field theory} allows a meaningful Fock 
expansion  of the interacting system state \cite{Brodsky} {(with the caveat of 
zero-modes)}. It has to be noted that in a field-theoretical 
framework, explicitly covariant, the constituent masses are off-shell and the four-momenta are conserved, but 
the interaction must be introduced perturbatively. On the contrary, in {a RHD} framework: (i) the explicit 
covariance is lost, (ii) the constituent masses are on-mass-shell and only three component of the momenta 
are conserved, but (iii) the Poincar\'e covariance fully holds, and (iv) the interaction can be introduced 
non perturbatively through the Bakamjian-Thomas construction of the Poincar\'e generators \cite{Baka}. 
This last feature is essential for a realistic description of nuclei.
 In this paper only the valence component of the LF wave function of the system is  considered and for the sake of definiteness we consider 
 the case of the three-nucleon systems, i.e., $^3$He and $^3$H.

The key quantity to be considered is the  { LF} spectral 
function, depending on {(i)  spin and  intrinsic  
momentum of the nucleon and (ii) } the removal energy of the two-nucleon spectator 
system
%{\Blue The unpolarized spectral 
%function of a nucleus  is essentially the probability distribution to find in the nucleus a nucleon %with a given value of the momentum, while the spectator system of the other (A - 1) nucleons has %a definite value for the energy} 
(for the definition of the non relativistic  
spin-dependent spectral function see, e.g., Ref. \cite{Ciofi}). With respect to previous attempts to describe DIS processes off 
$^3$He in a LF framework (see, e.g., the one in Ref. \cite{OSC}), in our approach a special care is devoted to the definition of 
the intrinsic LF variables of the problem, as well as to the spin degrees of freedom in the definition of the spin-dependent 
spectral function.
%through the Melosh rotations.
{In general, for an $A$-body system the spin-dependent spectral function yields 
 the probability distribution 
to find   a  constituent with a definite value of spin and   momentum, 
while the  $(A-1)$-constituent spectator system has a definite value of its mass. 
Such a distribution, properly convoluted  with the probe-nucleon elementary cross-section, leads to the description
of scattering processes off nuclei in 
impulse approximation. In this case,  the motion of the knocked 
out nucleon is free, while the spectator  system is fully interacting. 
Therefore one has to  relativistically describe a  final state} where the cluster separability {should be implemented}. 
As shown in Ref. \cite{KP}, this can be achieved {by adopting the tensor product of a 
plane wave for the knocked out constituent  and a fully interacting intrinsic 
state for the spectator  system, with given mass, all moving  {\em in their intrinsic reference frame}.
In order to build the spin-dependent spectral function, one needs to evaluate overlaps between the final state, previously described,}
and the ground state of the three-nucleon system. As a consequence 
a crucial part of the paper is devoted 
to carefully define
interacting and non interacting two- and three-body LF states,
{also providing the detailed link with the instant form counterparts. Notably, given the BT framework we have assumed, the instant form
states  in turn can be safely 
approximated by the corresponding 
non relativistic quantities, as explained in what follows. It should be pointed out that in order to }describe the needed states,  
three  
%different 
reference frames are considered: i) the laboratory 
frame of  the  fully interacting three-body system; ii) the intrinsic LF frame of three free particles;  iii) the intrinsic LF frame of a cluster 
of a free particle and an interacting two-particle subsystem.

%As a consequence,
{ Our} paper is organized as follows:
in Section \ref{lfk} the LF kinematics is  
summarized and in Section \ref{lfd23} the LF dynamics of two- 
and three-particle systems is briefly described, {and  
whenever} possible, use has been made of appendixes to
collect and discuss in detail the relevant formal results.
In {Section \ref{sf}} {the definition of the LF spin-dependent spectral function 
 in terms of the above mentioned overlaps, as well as}
 the LF momentum distributions and their sum rules are presented. 
  Conclusion and perspectives are discussed in Section \ref{concl}.

 \section{Light-front kinematics}
 \label{lfk}
{In this Section, for the sake of completeness and to establish the formalism, we   briefly review {the}
LF kinematics \cite{KP}.} 

{A generic LF four vector is  $v=(v^-,{\bf \tilde  v})$, with   
${\bf \tilde  v}=(v^+,{\bf v}_ \perp)$ and $v^\pm=v^0\pm v^3$; moreover 
the scalar product of two vectors ${\bf a}$ and ${\bf b}$ 
is given by }${\bf a}\cdot {\bf b} = {(a^-b^+ +a^+b^-)/ 2}-{\bf  a}_ \perp \cdot {\bf   b}_ \perp$.

Let us consider 
a system {of mass $M$} of $n$ on-mass-shell {\em interacting} particles of {mass $m_i$,} momenta $p_i ~~(i=1,...,n)$ 
and total momentum {$P$} 
{in the laboratory frame ($P^2= M^2$)}. 
%Since the particles are on mass shell, one has 
{ The minus components of the momenta are}
\be 
p^-_i={{ m^2_i+|{\bf p}_{i \perp }|^2 \over p^+_i}}
\label{meno}
\ee 
{ and the}
 following intrinsic  variables (invariant under a LF boost) can {be} introduced
\be 
\xi_i={p^+_i\over P^+}
 \quad \quad ,
%{ and }
 \quad \quad
{\bf k}_{i \perp }={\bf p}_{i \perp }- {p^+_i \over P^+} {\bf P}_{\perp}=
{\bf p}_{i \perp }-\xi_i {\bf P}_{\perp}
\label{cons1}   ~~~~~.
\ee
The conserved total LF-momentum of the system (a three-dimensional one!) is given by 
\be P^+=\sum_{i=1,n} p^+_i  \quad \quad , \quad \quad
{\bf P}_{\perp }=\sum_{i=1,n}{\bf p}_{i \perp } \quad \quad
\ee
and as a consequence one has
\be
\sum_i\xi_i=1  \quad \quad  , \quad \quad  \sum_i{\bf k}_{i \perp }=0\quad \quad .
\label{cons2}
\ee
One can complete the intrinsic variables adding the plus and  minus components
of the intrinsic momenta as follows: 
\be
k^+_i= \xi_i~M_0 
\nonu
k^-_i= {P^+ \over M_0}\left[p^-_i-2 {\bf p}_{i \perp }\cdot 
{{\bf P}_{\perp} \over P^+} +p^+_i \left({{\bf P}_{\perp} \over P^+}\right
)^2\right]=
{ m^2_i+|{\bf k}_{i \perp }|^2 \over k^+_i} \quad ,
\ee

where $M_0$ is the invariant (for LF boosts) free mass, given by 
\be
M^2_0=P^+~\sum_i{ m^2_i+|{\bf p}_{i \perp }|^2 \over p^+_i}
 -|{\bf P}_{\perp }|^2=\sum_i{ m^2_i+|{\bf k}_{i \perp }|^2 \over \xi_i} \quad .
\label{lf2}
\ee

Then, in a more compact form
\be
k_i^{\mu}= \left [B_{LF}^{-1}
\left ( {{\blf P}/M_0} \right )\right ]^{\mu}_{~\nu} p_i^{\nu} 
\ee
with
 $B_{LF}\left ( {{\blf P}/ M_0} \right )$  a LF boost to the intrinsic rest frame of the system
 of $n$ free particles of momenta $p_i$. 
  Such a  frame  is defined by a total LF-momentum ${\blf P}_{intr}\equiv \{\sum_i{k^+_i} = M_0, {\bf 0}_\perp\}$.

 Notice that
$k^2_i=p^2_i=m_i^2$, since in the LFHD the constituents are put on the
mass shell, as already mentioned. This feature, with  the nice  separation of the intrinsic motion from
the global one, as shown in Eqs. (\ref{lf2}) and (\ref{lf1}) (see below), make  
straightforward the  analogy with the non relativistic case.

Instead of the intrinsic variables $\xi_i$, one can introduce an alternative set of variables, namely
\be
k_{iz}= {1 \over 2}~\left[ k^+_i -k^-_i\right]= {M_0 \over 2}~\left[\xi_i-
{ m^2_i+|{\bf k}_{i \perp }|^2 \over M^2_0\xi_i}\right] 
\label{kzz}
\ee
that fulfill the following constraint (cf Eqs. (\ref{cons2}) and  (\ref{lf2}))
\be
\sum_{i=1,n}k_{iz}=0 \quad .
\ee
Then, one can equally well use the LF intrinsic variables, $\{k^+_i,{\bf k}_{i \perp }\}$, or the
Cartesian intrinsic variables, ${\bf k}_i$,  that fulfill
\be \sum_{i=1,n}{\bf k}_i=0 \quad .
\ee
To adopt the variables  ${\bf k}_i$ is useful for {making evident the analogy with the non relativistic 
framework, still remaining in  LFHD approach}.
In the case of free particles the intrinsic LF frame, defined by ${\blf P}_{intr}\equiv\{M_0,{\bf
0}_\perp\}$, can be also defined by ${\bf P}\equiv{\bf 0}$.
 {Let us recall that the bold character indicates a Cartesian vector, while the added tilde symbols indicates a LF three-vector.}
%, since $P^-_{free}=M_0$.

Because of the positivity of $\xi_i$, one can invert Eq. (\ref{kzz}) obtaining
\be
\xi_i={k_{iz}+\sqrt{m^2_i+|k_{iz}|^2+|{\bf k}_{i\perp}|^2} \over M_0}=
{k_{iz}+E_i \over M_0}\label{xi1} \quad ,
\ee
where $E_i = \sqrt{m^2_i +|{\bf k}_{i}|^2}$.
Then
\be
\sum_{i=1,n}E_i=M_0 \quad .
\ee
Let us stress that the minus component of the total momentum, $P^-$, is different from the free one
\cite{KP}
\be 
P^- = {{M^2 + {\bf{P}}^2_{\perp} \over P^+}} \ne \sum_{i=1,n} p^-_i=\sum_{i=1,n}
{ m^2_i+|{\bf p}_{i \perp }|^2 \over p^+_i}={1 \over P^+}\sum_{i=1,n}
{ m^2_i+|{\bf p}_{i \perp }|^2 \over \xi_i}=
P^-_{free} \quad .
\ee

In terms of the free mass, one can rewrite $P^-_{free}$ as follows
\be
P^-_{free}= {1 \over P^+} \left[ M^2_0 +|{\bf P}_{\perp}|^2\right] \quad .
\label{lf1}\ee

{For a particle of mass $m$,} {the LF spin, that has  the three components $s_{LF}^j$ in the particle
rest frame, yields  the Pauli-Lubanski four-vector in the reference where the
particle has LF-momentum $\blf p$, by applying  a proper LF boost,
$B_{LF}
\left ( {\blf p}/m \right )$ (see, e.g., Ref. \cite{PG} for a detailed discussion
of the LF spin). 
  On the other hand,  the canonical spin (instant form), $s_c^i$, is obtained {through}   a canonical boost,
  $B^{-1}_{c}
\left (  {\bf p}/m \right )$, {applied}
to the same  Pauli-Lubanski four-vector. Therefore, the relation between the 
two spins is given by 
\be
s_c^i=\left [B_{c}^{-1}
\left ( { {\bf p}/m} \right )\right ]^{i}_{~\nu} \left [B_{LF}
\left ( {{\blf p}/m} \right )\right ]^{\nu}_{~j} ~s_{LF}^j= 
 \left [ {\cal R}^\dagger_M ({\blf p}) \right ]^i_{j}~s_{LF}^j \quad ,
\ee
where ${\cal R}_M ({\blf p})$, called Melosh rotation {\cite{Melosh,Terent}}, is the 
rotation between the two rest  frames reachable through LF and canonical boosts,
respectively \cite{KP}.
 This rotation of spins implies the following relation between the plane wave 
 states of a particle with spin $s$ (notice that the squared spin does not
 depend on the chosen RHD form) in the  instant form and the LF one  
\be
|{\bf p} ;s\sigma \rangle_c =\sum_{\sigma'} ~D^{s}_{\sigma' \sigma}
\left[{\cal R}_M ({\blf p})\right]~|{\blf p};s \sigma'
\rangle_{LF} \quad ,
\label{melosh}
\ee
where $~D^{s}_{\sigma' \sigma}
\left[{\cal R}_M ({\blf p})\right]$ is the  Wigner function for a spin $s$. 
Within {\bf SL}(2C), the covering set of
the four-dimensional Poincar\'e group, the representation of the
 Melosh rotation for $s=1/2$, relevant in what follows, is a 
 $2\times 2$ matrix and  reads as follow 
\be
 D^{{1 \over 2}} [{\cal R}_M ({\blf k})]_{\sigma\sigma'}=~
\chi^\dagger_{\sigma}~{m+k^+-\imath 
{\bm \sigma} \cdot (\hat z \times
{\bf k}_{\perp}) \over \sqrt{\left ( m +k^+ \right )^2 +|{\bf k}_{\perp}|^2}}
~\chi_{\sigma'}=  _{LF}\langle {\blf k}; s \sigma | {\bf k}; s \sigma'\rangle_c \quad ,
\ee
where $\chi_{\sigma}$ is a two-dimensional spinor. The main feature 
of  LF rotations, $R_{LF}$, is given by the difference between  the corresponding 
Wigner rotations (that occurs when the  state $| {\blf k}; s \sigma'\rangle_{LF}$ has to be   transformed) and  the rotations
itself, differently from  the case of instant-form rotations $R_{IF}$  {(where $R_{IF}$ coincides with the associated Wigner
rotation)}
 \cite{KP,PG}. This 
  prevents the use of the  usual Clebsch-Gordan coefficients for 
  constructing the spin-spin and orbital-spin couplings within a LF framework,
and therefore one has to exploit the relation (\ref{melosh}) with the canonical spin.
 }
%in the frame moving with four-velocity $P^\mu/M_0$}. 

We adopt the following normalization for the LF
states $|{\blf p};s \sigma\rangle_{LF}$ 
\be
_{LF}\langle\sigma' s, {\blf p}'|{\blf p};s \sigma \rangle_{LF} =
2 p^+ (2\pi)^3~\delta^3({\blf p}'-{\blf p})~\sum_{\mu' \mu} ~
D^{s}_{ \sigma'\mu'}
\left[{\cal R}_M ({\blf p})\right]D^{s}_{\mu\sigma }
\left[{\cal R}^\dagger_M ({\blf p})\right] ~_c\langle\mu' s|s \mu\rangle_c=\nonu =
2 p^+ (2\pi)^3~\delta^3({\blf p}'-{\blf p})~\delta_{\sigma' \sigma}
\ee
and for the instant form states and spinors 
\be
\langle {\bf p}^\prime| {\bf  p}\rangle~=~2 E~ (2 \pi)^3~\delta( p^\prime_z - 
 p_z)\delta({\bf p}^\prime_\perp - 
{\bf p}_\perp)~~, \quad \quad \quad \quad
\bar u ~u=2 m,\quad \quad
u^\dagger u= 2 E
\label{pwnorm}\ee
 with $E({\bf  p})=\sqrt{m^2+|{\bf  p}|^2}$ and 
$\partial p^+/\partial p_z=1+p_z/p_0=p^+/p_0$.

  \section{Light-front dynamics for two- and three-particle systems}
\label{lfd23}
 In this Section a resum\'e of the main features of the BT construction, 
that allows  one to consistently include the interaction in the generators of the 
Poincar\'e group (see, e.g., \cite{KP}),
is presented. In particular, since   for defining   the LF   spectral function one needs  overlaps between the three-nucleon ground  state and  three-nucleon states composed by the tensor
product of a plane wave  for one of the particles and a two-body interacting 
state  for the spectator pair, we will focus on two- and three-body cases.

\subsection{Dynamics of two interacting particles} 

  In the case of a system of two identical particles, the LFHD leads to an Ansatz for the {two-body} mass
 operator able  to naturally embed a description 
 based on the Schr\"{o}dinger
 equation into a Poincar\'e-covariant framework (see, e.g.
 \cite{LPSD,LPS2,LPS3} for an application).
 
  {By eliminating the longitudinal LF variable $\xi$ in favor of   the third Cartesian component of the intrinsic momentum
\be
k_z=k_{1z}=M_0(1,2)~(\xi -{1\over 2})  \quad ,
\ee
where $M^2_0(1,2)$ is given by
\be
M^2_0(1,2)={m^2 +|{\bf k}_\perp|^2 \over \xi (1-\xi)}=4~\left[E(\bf  k)\right]^2=
4~(m^2 +|{\bf k}|^2) \quad ,
\label{M02b}
\ee
one can show the  formal equivalence between a non relativistic
description and a LF one.} 
Moreover, one has 
\be k^+_1 = \xi~M_0(1,2) =k^+
\quad \quad ,  \quad \quad
k^+_2 = (1-\xi)~M_0(1,2) = M_0(1,2) - k^+ \quad .
\label{intrp}
\ee

 The two-body Hamiltonian, with an interaction that depends upon
 intrinsic variables and fulfills the correct  invariance properties under rotations and translations,
  leads to a square mass
 operator suitable for  a  Bakamijan-Thomas (BT) construction of the Poincar\'e generators \cite{Baka}.  {This construction gives a simple way to 
 introduce the interaction in the generators, while satisfying the correct commutation rules. } As a matter of fact, within the BT framework 
 the two-body mass equation  can be written  as follows {(see, e.g.,  \cite{KP,coester,Lev1})}
 {\be
 \langle \sigma_1,\sigma_2;  \tau_1, \tau_2; {\bf k}|\left[M^2_0(1,2) +U(|{\bf k}|)\right]~|j,j_z;
\epsilon_{int},\alpha; T T_z 
 \rangle = M^2 ~\langle \sigma_1,\sigma_2;  \tau_1, \tau_2; {\bf k}|j,j_z;
\epsilon_{int},\alpha; T T_z \rangle \nonu \langle \sigma_1,\sigma_2;  \tau_1, \tau_2; {\bf k}|\left[4m^2 + 4|{\bf k}|^2 
+U(|{\bf k})|\right]~|j,j_z;
\epsilon_{int},\alpha; T T_z \rangle =M^2 ~\langle \sigma_1,\sigma_2;  \tau_1, \tau_2; {\bf k}|j,j_z;
\epsilon_{int},\alpha; T T_z \rangle
\nonu
\langle \sigma_1,\sigma_2;  \tau_1, \tau_2; {\bf k}|\left[{|{\bf k}|^2\over m} +V(|{\bf k}|)\right]
~
|j,j_z; \epsilon_{int},\alpha; T T_z \rangle =
%\left [{M^2 -4m^2\over 4 m}\right]
\epsilon_{int}~\langle \sigma_1,\sigma_2;  \tau_1, \tau_2; {\bf k}|j,j_z; \epsilon_{int},\alpha; T T_z \rangle  \quad ,
\label{eigen2} 
\ee}
where {$V=U/(4m)$} and 
\be
\epsilon_{int}={M^2 -4m^2\over 4 m} \quad .
\label{eigen3}
\ee
 In the last line of Eq. (\ref{eigen2}) one formally recovers the Schr{\"o}dinger 
equation for a two-body  {intrinsic eigenstate (that does not depend upon the chosen RHD)} of angular momentum $(j,j_z)$, intrinsic energy $\epsilon_{int}$
(negative for bound states and positive for the scattering ones) and isospin $(T,T_z)$. The symbol $\alpha$ represents the quantum numbers needed 
to completely define the state of the system.
For the bound state (the deuteron in our case) one has $M=2m-B$, and then

\be 
\epsilon_{int} = -B+{B^2 \over 4m}\sim -B \quad \quad  ,
\label{eigen4}\ee
given the  small binding energy of the deuteron with respect to its mass.
For the scattering states, one has $M^2=s$, with $s$  one of the Mandelstam 
variables,  and asymptotically 
$M^2=4m^2 + 4|{\bf t}|^2$ with ${\bf t}$ the asymptotic Cartesian momentum in
the intrinsic frame. Then, one can write
\be 
\epsilon_{int}={M^2 -4m^2\over 4 m}=
{|{\bf t}|^2\over m} \quad .
\label{eigen5}
\ee
Therefore the intrinsic eigenstates of Eq. (\ref{eigen2}) 
(i.e. of a Poincar\'e covariant mass operator)
can be safely identified with the usual non relativistic two-body 
eigenstates \cite{LPS2,LPS3} (only for bound states one disregards terms {$O(B/(4m))$}) 
{and  the overlap 
$\langle \sigma_1,\sigma_2;  \tau_1, \tau_2; {\bf k}|j,j_z;\epsilon_{int},
\alpha ; T T_z \rangle$,  that contains  canonical spins,}
  with  its {\em non relativistic  counterpart}.

As discussed in Appendix \ref{appa}, the normalized LF two-body wave function  is
\be
 _{LF}\langle \sigma_1,\sigma_2;  \tau_1,\tau_2;{\blf k},{\blf P}'
 |{\blf P};j,j_z;
 \epsilon_{int},\alpha; T T_z \rangle_{{LF}}  
 =2~P^+~
 (2\pi)^3~\delta ^3({\blf P}'  -{\blf P})~ \sqrt {(2\pi)^3 E({\bf k}) }
% \left[k^+~(2\pi)^3{\partial k_z \over \partial k^+}\right]^{1/2} 
 ~\times  \nonu
 \sum_{\sigma'_1,\sigma'_2}
~ D^{{1 \over 2}} [{\cal R}_M({\blf k} )]_{\sigma_1\sigma'_1}~
D^{{1 \over 2}} [{\cal R}_M (-{\blf k} )]_{\sigma_2\sigma'_2}~
~\langle \sigma'_1,\sigma'_2;  \tau_1,  
\tau_2; {\bf k}|j,j_z;
\epsilon_{int},\alpha; T T_z 
 \rangle~~~~,
\label{lfwf22}
\ee 
where we define (cf Eq. (\ref{intrp}))
\be
-{\blf k}\equiv((M_0 - k^+),-{\bf k}_\perp) \quad \quad  .
\ee
It has to be emphasized that    in the  intrinsic two-body wave function 
$\langle \sigma'_1,\sigma'_2;  \tau_1,  \tau_2; {\bf k}|j,j_z; \epsilon_{int},\alpha; T T_z  \rangle $  
the  canonical spins can be composed with the orbital angular momenta 
by using the familiar Clebsh-Gordan
 coefficients.
 The 
 state $|{\bf k}\rangle$ (with Cartesian variables) is normalized as follows
  \be
 \langle {\bf k'}| {\bf k}\rangle= \delta ({\bf k'} - {\bf k}) \quad \quad .
  \ee
  {Notice the difference with Eq. (\ref{pwnorm}).}
Furthermore, for the two-body interacting case the LF completeness reads (see Eq. (\ref{tbcompl}))
\be
\hspace{0.5cm} \int {d{\blf P} \over 2 P^+ (2 \pi)^3} \sum_{j,j_z\alpha} \sum_{T T_z}
 \sumint {\lambda(t) ~
 dt
 }
|{\blf P};j,j_z; \epsilon_{int},\alpha; T T_z \rangle_{LF} ~
_{LF}\langle T_z T;\alpha,\epsilon_{int};j_z,j
;{\blf P}|= {\bf I} \quad ,
%\nonu
\label{tbcompl2}
\ee 
where the symbol $\int \! \! \! \!\! \!\sum$ means a sum over the
bound states of the  pair (namely the deuteron in the present case) 
and an integration over the 
continuum. Notice the choice of the Cartesian $t$ momentum to label the intrinsic energy. 
The quantity $\lambda(t)$ is the t-density of the 
two-body states ($\lambda(t)= 1$ for the bound states and $\lambda(t)= t^2 $ for the continuum).
Such a completeness follows from the one fulfilled by the
eigensolutions of Eq. (\ref{eigen2}), i.e.
\be
\sum_{j,j_z\alpha} \sum_{T T_z}\sumint {\lambda(t) ~ dt}
\langle{\bf k}' |j,j_z; \epsilon_{int},\alpha; T T_z \rangle ~
\langle T_z T;\alpha,\epsilon_{int};j_z,j|{\bf k}\rangle=
\delta^3({\bf k}'-{\bf k}) \quad \quad .
\label{nrcompl1}
\ee

\subsection{Three interacting particle systems} 
In order to have a Poincar\'e covariant description of an interacting system,
 like the $^3$He nucleus, it seems appropriate to adopt the LFHD framework,
  combined with a Bakamjian-Thomas (BT) construction \cite{Baka} of the Poincar\'e
 generators. With a suitable  Ansatz for the interaction (see e.g. \cite{KP,Lev1}), the mass operator is
 \be
 M(1,2,3)=M_0(1,2,3) + {\cal V}(1,2,3)=
 \sum _{i=1,3}\sqrt{m_i^2+|{\bf k}_i}|^2+{\cal V}({\bf k}_i\cdot{\bf k}_j) \quad ,
 \label{M3}
 \ee
 where ${\bf k}_i$ are the intrinsic momenta defined in Sect. \ref{lfk}, and the interaction ${\cal V}$ is 
 invariant for rotations and translations. {The ground state can be written as  the product of a plane wave describing the global
 motion with LF momentum ${\blf P}$ times eigenvectors of the three-body mass operator in Eq. (\ref{M3}). It reads
 \be
 |{\blf P};  j,j_z;
 \epsilon^3_{int},\Pi; {1 \over 2}, T_z \rangle_{LF} \quad ,
 \label{3bwf}
 \ee
 where  {$\epsilon^3_{int}= M_3 - 3m$} is the energy, $j$ the total angular momentum, $1/2$ the isospin of the {system}
 and $\Pi$  the parity.} {From now on, we assume that the three particles have the same mass.}
 
 When applications like DIS or SIDIS processes are concerned, the issue of 
 macrocausality has to be considered, i.e.   if the subsystems which compose a
  system are brought far apart, the Poincar\'e generators 
 of the system have to become the sum of the Poincar\'e
 generators corresponding to the subsystems in which the system is asymptotically separated.
 It is important to notice that the packing operators  
 \cite{soko,KP}, 
 that make it possible to include the
 macrocausality, are  not considered in the present approximation for the 
 description of the bound state. However, we  implement macrocausality in the
  tensor product of a plane wave for the knocked out constituent times a 
  fully interacting intrinsic state for the spectator pair. This tensor
   product 
  {is needed for the definition of the LF Spectral Function, as shown below.}
%   is not relevant for our present studies, that are 
% mainly devoted to the dynamical description of the ground state. Indeed, 
% one can argue that the overlap
% formalism we have to introduce for obtaining the LF Spectral Function should 
 %help for implementing the macrocausality, when needed.

%In the intrinsic frame of the three-particle system,  instead of using
% $\{\xi_i,{\bf k}_{i\perp}\}$ one can use
%$\{k_{iz},{\bf k}_{i\perp}\}$.
%Then, one can introduce a set of  Jacobi momenta (Cartesian momenta) 
%and conjugate coordinates as follows
%\be {\bf p }= { 2 \over 3} \left (
%{\bf k}_1 - {{\bf k}_2 + {\bf k}_3 \over 2} \right )
%= {\bf k}_1 \quad \quad \rhobf= \left (
%{\bf r}_1 - {{\bf r}_2 + {\bf r}_3 \over 2} \right )
%\nonu {\bf k} = { 1\over 2} \left({\bf k}_2 -{\bf k}_3\right ) = 
%{\bf k}_2 + { {\bf k}_1\over 2} \quad \quad ~{\bf r}= \left({\bf r}_2 -{\bf r}_3 
%\right ) 
%\label{Jacobi}
%\ee 
%Notice that the Jacobian of the transformation is 1 for both momenta and coordinates.

In a given frame, the LF three-body wave function can be expressed in terms of
the intrinsic wave function, with canonical spins. {Therefore}, as in the two-body case, one can
approximate such an intrinsic wave function by 
the corresponding non relativistic wave function, after checking that the non relativistic 
Schr{\"o}dinger operator can be properly identified with a BT mass operator. 
 Then the key point {for actual calculations} is the approximation $M(1,2,3)\sim M_{NR}(1,2,3)$, 
which is based on an appropriate definition of the interaction ${\cal V}$. This approximation is allowed since  
\be
M_{NR}(1,2,3)=3m + \sum_{i=1,3} {k^2_i / 2m}
+V^{NR}_{12}+V^{NR}_{23}+V^{NR}_{31}+V^{NR}_{123}
\ee
 fulfills rotational 
and translational invariance, {namely the general properties for making
a mass operator acceptable
 as a BT mass operator. {As a matter of fact, those properties are just the
 ones satisfied by the  non relativistic nuclear interactions 
 that give  an accurate description of {two- and 
 three-nucleon} data  (see, e.g., \cite{AV18,CPW}). 
  An early
 investigation  of the electromagnetic tri-nucleon systems, within the above
 illustrated approach  and
 using the refined non relativistic ground states of Ref. 
 \cite{pisa}, can be found in  Ref.
 \cite{Baroncini}.

%The normalized LF three-body wave function in a generic frame is given by (let us recall
%that the LF boosts are kinematical)
%\be
% _{LF}\langle \sigma_1,\sigma_2,\sigma_3; T_{23}, \tau_{23},  
%\tau;{\blf p}_1,{\blf p}_2,{\blf p}_3|j,j_z;
%%
%(2\pi)^3~\delta ^3({\blf p}_1 + {\blf p}_2 + {\blf p}_3 -{\blf P})
% ~\nonu \times ~\sum_{\sigma'_1}
%\sum_{\sigma'_2}
%\sum_{\sigma'_3}~
 %D^{{1 \over 2}} [{\cal R}_M^\dagger ({\blf k} _1)]_{\sigma_1\sigma'_1}
%~
%D^{{1 \over 2}} [{\cal R}_M^\dagger ({\blf k} _2)]_{\sigma_2\sigma'_2}~ D^{{1 \over 2}}
 %[{\cal R}_M^\dagger( {\blf k} _3)]_{\sigma_3\sigma'_3}
%~\nonu \times ~~
%\sqrt{4(2\pi)^6{E_1 E_2 E_3 \over  M_0(1,2,3) }}
%\langle \sigma'_1,\sigma'_2,\sigma'_3; T_{23}, \tau_{23},  
%\tau; {\bf p},{\bf k}|j,j_z;
%\epsilon^3_{int},\Pi; {1\over 2} T_z  \rangle  
%\label{lfov}\ee 
%where  $T_{23}$ is the isospin of the pair $(2,3)$, $\tau_{23}$ the
%corresponding third component,  $\tau$ the third component of particle $1$ isospin, while
%$j$, $j_z$, $\epsilon^3_{int}$, $\Pi$ and $T_z$ are the total angular momentum, its third %component, 
%the intrinsic energy,
 %the parity and the isospin third component  of the three-body system, respectively. 

\subsubsection{Non-symmetric intrinsic variables}
To define the LF   spectral function one needs the overlaps between the ground state  of  the three-body system and the states {composed by the
tensor product}  of a free nucleon 
and a  fully  interacting two-body system. Therefore proper variables, suited to  
describe  these {states}, have to be introduced. 
Instead of the symmetric intrinsic variables ${\blf k}_i$ ($i=1,2,3$) that 
refer to the three particles moving in the three-body intrinsic frame, {it is
more suitable to introduce non symmetric variables.}
Let us consider the intrinsic variable ${\blf k}_j$ for particle $j$ and the
 intrinsic variables for the internal motion of the spectator pair.
For the sake of concreteness, let us take $j=1$ and   focus on  the 
 kinematics of the (2,3) pair, 
 that globally moves in the three-body intrinsic frame with total LF momentum 
($K^+_{23},K_{23\perp}$). 
A set of intrinsic variables for the internal motion of the (2,3) pair can be 
defined {as follows}
\be
  \eta= {k^+_2 \over k^+_2+k^+_3}={\xi_2 \over (\xi_2 +\xi_3)}={\xi_2\over
  1-\xi_1}={p^+_2 \over p^+_2+p^+_3}
 \nonu {\bf k}_{23\perp} ={\bf k}_{2\perp}-\eta ({\bf k}_{2\perp}+{\bf k}_{3\perp})=
  {\bf k}_{2\perp}+\eta {\bf k}_{1\perp}
%  ={\bf p}_{2\perp}-\eta ({\bf p}_{2\perp}+{\bf p}_{3\perp})
  \nonu k^+_{23}=~\eta M_{23}
 \nonu k_{23z}= M_{23}~(\eta-{1 \over 2}) \quad ,
 \label{nonsymv}
 \ee
 where {$k^+_{i}=\sqrt{m^2+|{\bf
k}_{i}|^2} +k_{iz}$ and  } $M_{23}$ is the free mass for the (2,3) pair, defined as in  Eq. (\ref{M02b}),
\be
M^2_{23}={m^2+|{\bf k}_{23\perp}|^2 \over \eta (1-\eta)}=
\left [ 2 \sqrt{m^2+|{\bf k}_{23}|^2}\right]^2 \quad .
%= 
\label{clus3}
\ee

 Furthermore, the total LF  momentum of the free (2,3) pair in the laboratory frame is
 \be
P_{23}^+= p_2^+ + p_3^+ 
\nonu
{\bf P}_{23\perp}= {\bf p}_{2\perp}+{\bf p}_{3\perp} 
\label{nonsymv1} \quad \quad ,
\ee 
while in the intrinsic three-body frame the total LF momentum is
 \be
K_{23}^+= k_2^+ + k_3^+ 
\nonu
{\bf K}_{23\perp}= {\bf k}_{2\perp}+{\bf k}_{3\perp} = - {\bf k}_{1\perp} \quad .
\label{nonsymv2}
\ee

In terms of the non-symmetric intrinsic variables, the free mass of the three-particle system can be written as follows
\be
\hspace{-.6cm} M_0(1,2,3) =
 \sum _{i=1,3}\sqrt{m^2+|{\bf k}_i}|^2=
 \sqrt{m^2+|{\bf k}_1}|^2 +  \sqrt{M_{23}^2+|{\bf k}_1}|^2 =
 {m^2+|{\bf k}_{1\perp}|^2 \over k^+_{1} }+ {M_{23}^2+|{\bf k}_{1\perp}|^2 \over K_{23}^+} \quad .
 \label{freens}
 \ee
 
{ Then one has
\be
{m^2+|{\bf k}_{2\perp}|^2\over k^+_2}
+{m^2+|{\bf k}_{3\perp}|^2\over k^+_3}= {1 \over k^+_2+k^+_3} \left[M^2_{23} +
|{\bf k}_{2\perp}+{\bf k}_{3\perp}|^2\right] \quad \quad ,
\ee
and therefore
\be
M^2_{23} = {{m^2+|{\bf k}_{2\perp}|^2\over \eta}
+{m^2+|{\bf k}_{3\perp}|^2\over (1-\eta)}-
|{\bf k}_{1\perp}|^2 } ~~~ .
\ee
}

\subsubsection{Three-body light-front wave function with non-symmetric intrinsic variables}

For the fully interacting case, i.e. ${\cal V}(1,2,3)\ne 0$, 
the  three-body LF wave function, {can be  expressed through (i) 
the non-symmetric intrinsic variables $\{{\blf k}_1,{\blf k}_{23}\}$ introduced
in the previous subsection, instead of
using the three-body standard Jacobi coordinates {(defined through 
$\tilde k_1, \tilde k_2, \tilde k_3$)},   and (ii) canonical spins in
the reference frame where $P^+=M_0(1,2,3)$.
Therefore, by repeating analogous steps as in the two-body case 
(cf Eq. (\ref{lfwf22})) one has}
\be
 _{LF}\langle \sigma_1,\sigma_2,\sigma_3;   \tau_1,\tau_2,\tau_3; 
{\blf k}_1,{\blf k}_{23},{\blf P}'|{\blf P};  j,j_z;
 \epsilon^3_{int},\Pi; {1 \over 2}, T_z \rangle_{LF}  
 =~2~P^+~
 (2\pi)^3~\delta ^3({\blf P}'  -{\blf P})
 \nonu \times~ \sum_{\sigma'_1}
\sum_{\sigma'_2}
\sum_{\sigma'_3}~
D^{{1 \over 2}} [{\cal R}_M ({\blf k}_1 )]_{\sigma_1\sigma'_1}~
D^{{1 \over 2}} [{\cal R}_M ({\blf k}_{2} )]_{\sigma_2\sigma'_2}~ 
D^{{1 \over 2}}
 [{\cal R}_M({\blf k}_{3} )]_{\sigma_3\sigma'_3}
~\nonu \times ~~\sqrt{{ (2\pi)^6~2E_1 E_{23} M_{23}\over  2 M_0(1,2,3) }} ~
\langle \sigma'_1,\sigma'_2,\sigma'_3;  \tau_1,\tau_2,\tau_3; {\bf k}_1, {\bf k}_{23}|j,j_z;
\epsilon^3_{int},\Pi; {1\over 2} T_z 
 \rangle  \quad \quad  ,
\label{lfwf4}
\ee
where  $E_{23} = \sqrt{M_{23}^2+|{\bf k}_1}|^2 $ and $M_{23} = 
[m^2 +  |{\bf k}_{23\perp}|^2 + (k^+_{23})^2]/k^+_{23}$. { The LF variables 
${\blf k}_{2}$ and ${\blf k}_{3}$ can be easily obtained from ${\blf k}_{1}$ 
 and ${\blf k}_{23} $. Indeed one has {(i) $\eta= k^+_{23}/M_{23}$, (ii)
 ${\bf k}_{2\perp} = {\bf k}_{23\perp}  - \eta {\bf k}_{1\perp}$, (iii)
%and 
%$M_0(1,2,3) =  \sqrt{m^2+|{\bf k}_1}|^2 +  \sqrt{M_{23}^2+|{\bf k}_1}|^2 $. Then 
%$k_{23}^+ = \eta ~ [M_0(1,2,3)  - k_1^+$]. Then 
${\bf k}_{3\perp} = -{\bf k}_{1\perp} - {\bf k}_{2\perp}$, (iv) $k_2^+ + k_3^+= M_0(1,2,3) - k_1^+ 
 $ (cf Eq. (\ref{freens})),  (v) $k_2^+=\eta~(k_2^+ +k_3^+)$ and 
 (vi)  $k_3^+= M_0(1,2,3) - k_1^+  - k_2^+$.}

{In  Eq. (\ref{lfwf4}), the intrinsic  wave function with canonical spins} 
$\langle \sigma'_1,\sigma'_2,\sigma'_3;  \tau_1,\tau_2,\tau_3; {\bf k}_1, 
{\bf k}_{23}|j,j_z;\epsilon^3_{int},\Pi; {1\over 2} T_z  \rangle  $  is the eigensolution of the mass operator $M(1,2,3)$ of Eq. (\ref{M3}), that in actual calculation can be approximated by the non relativistic Hamiltonian operator (since, we repeat, the symmetry requirements are the same). 
As shown in Appendix \ref{appb} (see Eq. (\ref{normhe3})), the factors in Eq. (\ref{lfwf4}) allow one to recover the normalization for the intrinsic
 part of the three-body 
bound state 
%{\Red CUT  \be
%\langle  {1 \over 2}, T_z ; \epsilon^3_{int},\Pi;  j,j_z |
%  j,j_z; \epsilon^3_{int},\Pi; {1 \over 2}, T_z \rangle = 1
%  \label{norm5}
%\ee}
according to
\be
\sum_{\tau_1,\tau_2,\tau_3}\sum_{\sigma_1,\sigma_2,\sigma_3}
\int d{\bf k}_1 \int d{\bf k}_{23}~|\langle \sigma_1,\sigma_2,\sigma_3; \tau_1,\tau_2,\tau_3; {\bf k}_1,
{\bf k}_{23}|j,j_z;
\epsilon^3_{int},\Pi; {1\over 2} T_z \rangle |^2=1 \quad \quad ,
\label{norm3}
\ee
like in the non relativistic case.

\subsubsection{Free-mass  and intrinsic reference frame for the (1,23) cluster}

Because of our interest  in constructing the overlap between the three-nucleon
ground state and a state where only the pair $(2,3)$ is interacting, while the
third nucleon is free, {in what follows we investigate the corresponding mass operator, whose eigenstates
are the tensor product we have already mentioned.}

By using the intrinsic variables $\{\xi_1,{\bf k}_{1\perp}\}$,
 %and the mass eigenvalue, $M_S$, of the interacting (2,3) pair,
 {one {can} introduce  the 
  squared free-mass, ${\cal{M}}^2_0(1,23)$,  for the {cluster} $(1,23)$, 
 when the mass
eigenvalue of the interacting $(2,3)$ pair is $M_S$ 
\be
%M^2_0(1,2,3)={m^2+|{\bf k}_{1\perp}|^2\over \xi_1}+{m^2+|{\bf k}_{2\perp}|^2\over \xi_2}
%+{m^2+|{\bf k}_{3\perp}|^2\over \xi_3}=\nonu
{\cal{M}}^2_0(1,23)={m^2+|{\bf k}_{1\perp}|^2\over \xi_1}
+{M^2_S+|{\bf k}_{1\perp}|^2\over (1-\xi_1)} \quad \quad .
\label{clus1}
\ee
The intrinsic frame of the cluster $(1,23)$ is defined by 
${\blf P}_{int}(1,23) \equiv \{{\cal M}_0, {\bf 0}_\perp\}$. 
In this frame, 
{the LF} momentum of the nucleon $1$  is given by}
\be
\kappa_1^+=\xi_1{\cal{M}}_0(1,23)
\nonu
{\bm  \kappa}_{1\perp}= {\bf p}_{1\perp}-\xi_1{\bf P}_{\perp}={\bf k}_{1\perp} \quad ,
\label{nint}
\ee
{while  the  {$z$} Cartesian component reads} (see Eq. (\ref{kzz}))
\be
\kappa_{1z}= {1 \over 2}~\left[ \kappa^+_1 - \kappa^-_1\right]= {{\cal{M}}_0(1,23) \over 2}~
\left[\xi_1-
{ m^2_1+|{\bm \kappa}_{ 1 \perp }|^2 \over {\cal{M}}_0(1,23)^2 ~ \xi_1}\right]   \quad \quad .
\label{kz}
\ee
As a consequence one has
\be
{\cal{M}}_0(1,23)=E({\bf {\kappa}}_1)+E_{S}
\label{clus2}
\ee
with $E({\bf {\kappa}}_1) = \sqrt{m^2 +|{\bm {\kappa}}_1|^2}$ and 
$E_{S} = \sqrt{M^2_S +|{\bm {\kappa}}_1|^2}$.

The total momentum of the (2,3) pair in the same frame is
\be
K_S^+=(1-\xi_1) ~ {\cal{M}}_0(1,23)
\nonu
{{\bf  K}_{S\perp}=  - {\bm \kappa}_{1\perp}=- {\bf k}_{1\perp}={\bf k}_{2\perp}+{\bf
k}_{3\perp}}
\nonu
K_{Sz} = - \kappa_{1z}
\nonu
K^-_{ S on} = {M^2_S+|{\bf k}_{1\perp}|^2\over K^{+}_S} \quad .
\ee

Summarizing the pair $(2,3)$, with {internal} variables  
$\{\eta,{\bf k}_{23\perp}\}$ and mass eigenvalue $M_S$ 
{(cf Eqs. (\ref{eigen2}), (\ref{nonsymv}) ),
}
 is moving with LF momentum 
${\blf K}_S $
in the intrinsic frame of the 
{\em three-particle cluster} (1,23). 

{It should be pointed out that  the intrinsic frame for 
the three-body system (1,2,3) and  the intrinsic frame of the (1,23) 
cluster are related by a proper  longitudinal LF boost that makes the change
$P^+_{int}(1,23)={\cal M}_0(1,23) \to P^+_{int}(1,2,3)={M_0(1,2,3)}$.} 

\subsubsection{Non-symmetric basis for three interacting particle systems}

  In the $1+(23)$ cluster  only the interaction  $U_{23}$ between particles 
2 and 3 is active; then one can introduce a three-body state
 given by the tensor product of an eigenstate of the total LF momentum, 
 ${\blf P}$, times  {{\em the intrinsic state of the cluster with a given mass for the interacting pair}.
   In turn, such an intrinsic state, that fulfills the
 macrocausality \cite{KP}, is given by the tensor product of 
 a plane wave for particle 1 with LF momentum $ \tilde{\bm \kappa}_1$},
 times the fully interacting  state of {the} pair corresponding to the given energy eigenvalue.  Therefore, one can write}
 \be
|{\blf P}; \tilde{\bm \kappa}_1 \sigma_1 \tau_1; 
j_{23} j_{23z} \epsilon_{23},\alpha;T_{23},\tau_{23}\rangle_{LF} \quad ,
\label{1e23}
\ee
 
%{\Red 
%\be= \sqrt{{\cal
%M}_0\over P^+}~ 
%U\Bigl[B_{LF}({\blf P})\Bigr]~|{\blf P}_{int}; \tilde{\bm \kappa}_1 \sigma_1 \tau_1; 
%j_{23} j_{23z} \epsilon_{23},\alpha;T_{23},\tau_{23}\rangle_{LF}
 %\quad \quad ,
%\ee
%{\Red where ${\blf P}_{int}\equiv \{{\cal M}_0(1,23), {\bf 0}_\perp\}$,
%$U\Bigl[B_{LF}({\blf P})\Bigr]$ the infinite-dimensional unitary representation
%of the needed LF boost and $|{\blf P}_{int}; \tilde{\bm \kappa}_1 \sigma_1 \tau_1; 
%j_{23} j_{23z} \epsilon_{23},\alpha;T_{23},\tau_{23}\rangle_{LF}$  }}
which is an eigenstate of the mass operator 

\be
M'(1,23)=E({\bf {\kappa}}_1)+\sqrt{M^{2}_{23}(|{\bf k}_{23}|)+ U_{23} + | \tilde{\bm \kappa}_1|^2} =
E({\bf {\kappa}}_1)+\sqrt{M^{^* 2}_{23}(|{\bf k}_{23}|)+ | \tilde{\bm \kappa}_1|^2} \quad \quad ,
\label{mass3}
\ee
with eigenvalue ${\cal{M}}_0(1,23)=E({\bf {\kappa}}_1)+E_{S}$ 
($E_S=\sqrt{M^2_S +|{\bm {\kappa}}_1|^2}$).
The {operator} $M^{^* 2}_{23}(|{\bf k}_{23}|)=M^{ 2}_{23}(|{\bf k}_{23}|)+ U_{23}(|{\bf k}_{23}|)$ 
is the square of the intrinsic mass operator of the
interacting (2,3) pair, with eigenvalue $M_S^2 = 4(m^2 + m \epsilon_{23})$ 
(see Eq. (\ref{eigen2})).

The set of eigenstates (\ref{1e23}) is complete with the following completeness relation
\be
 \int {d{\blf  P}\over 2P^+(2\pi)^3}
~\sum_{T_{23} \tau_{23}} ~
\sumint {\lambda(t)~d{  t}
}  
\sum_{j_{23}j_{z23} \alpha} ~
\sum_{\sigma_1 \tau_1}
\int {d\tilde{\bm \kappa}_{1}\over 2 \kappa^+_1 (2\pi)^3}
 ~\nonu \times ~
~  
|{\blf P}; \tilde{\bm \kappa}_1 \sigma_1 \tau_1; j_{23},
j_{z23}; \epsilon_{23},\alpha; T_{23}, \tau_{23} \rangle_{LF}~ 
_{LF} \langle T_{23}, \tau_{23}; \alpha,\epsilon_{23};j_{z23},j_{23}; 
\tau_1 \sigma_1 \tilde{\bm \kappa}_1;{\blf P}| = ~ \bf I   \quad \quad .
\label{complns}
\ee
{Since it will play a relevant role for a proper definition of the LF spectral function,}  let us consider the    overlap 
{between the eigenstates (\ref{1e23}) and {the product of plane waves for (i) the total LF momentum {$P'$ for a  system of three free  particles, (ii) 
the LF momentum of particle $ 1$, ${\blf k}'_1$, in the intrinsic frame of {\em the three free particles %system
}  and (iii) the LF momentum, ${\blf k}'_{23}$, for  the intrinsic motion of the free subsystem $(2,3)$}.} One has
\be
~_{LF}\langle \sigma'_1,
\sigma'_2,\sigma'_3; \tau'_1,  \tau'_2, \tau'_3 ;
{\blf P}',{\blf k}'_1,{\blf k}'_{23}|{\blf P}; \tilde{\bm \kappa}_1 \sigma_1 \tau_1; j_{23},
j_{23z};
 \epsilon_{23},\alpha; T_{23}, \tau_{23} \rangle_{LF} =\nonu
 =~2~P^+~
 (2\pi)^3~\delta ^3({\blf P}' -{\blf P})
 %~ \delta_{\sigma_1\sigma'_1}
 ~\delta_{\tau_1\tau'_1}~
 ~_{LF}\langle{\sigma'_1{\blf k}'_{1} }| { \tilde{\bm \kappa}_1} \sigma_1\rangle_{LF}~
%{(2\pi)^3~2 k^{'+}_1 }~ \delta^3({\blf k}'_1 -{\blf k}_1^{(a)})~
%\sqrt{\kappa_1^+ E'_{23}\over k^{'+}_1 E_S}
~_{LF}\langle
\sigma'_2,\sigma'_3; \tau'_2, \tau'_3 ;
{\blf k}'_{23}| j_{23},j_{23z};
 \epsilon_{23},\alpha; T_{23}, \tau_{23} \rangle_{LF} =
 \nonu
 ~ \nonu
  =
  ~2~P^+~
 (2\pi)^3~\delta ^3({\blf P}' -{\blf P})
 %~ \delta_{\sigma_1\sigma'_1}
 ~\delta_{\tau_1\tau'_1}~
% ~\sum_{\sigma'\sigma''}D^{{1 \over 2}} [{\cal R}^\dagger_M(\tilde{\bm \kappa}_{1} )]_{\sigma''\sigma_1}
%D^{{1 \over 2}} [{\cal R}_M({\blf k}'_1)]_{\sigma'_1\sigma'}~\delta_{\sigma'\sigma''}~
{ \delta_{\sigma_1\sigma'_1}}~\nonu \times ~
{(2\pi)^3~2 k^{'+}_1 }~ \delta^3({\blf k}'_1 -{\blf k}_1^{(a)}) ~ \sqrt{\kappa_1^+ E'_{23}\over k^{'+}_1 E_S} ~
~\sqrt{(2\pi)^3{E'_{23} ~ M'_{23}\over 2 {{M}}'_0(1,2,3)}}~
\nonu \times ~
\sum_{\sigma_2}
\sum_{\sigma_3} ~
~
D^{{1 \over 2}} [{\cal R}_M({ {\blf k}'_{23}})]_{\sigma'_2\sigma_2}~ 
D^{{1 \over 2}}
 [{\cal R}_M({ {-\blf k}'_{23}} )]_{\sigma'_3\sigma_3}
~
\langle \sigma_2,\sigma_3; \tau'_2, \tau'_3 ; {\bf k}'_{23}|j_{23},j_{23z};
\epsilon_{23},\alpha;   T_{23}, \tau_{23} \rangle  ~\quad ,
\label{lfwf3}
\ee
 where 
 {$E'_{23}=\sqrt{{M'}^2_{23} + {\bf k}'^2_1}$}, 
 $M'_{23} = [m^2 +  |{\bf k}'_{23\perp}|^2 + (k'^+_{23})^2]/k'^+_{23}$~~, 
  \be
 \hspace{-.5cm} M'_0(1,2,3) =
 \sqrt{m^2+|{\bf k}'_1}|^2 +  \sqrt{{M'}_{23}^2+|{\bf k}'_1}|^2  \quad \quad  ,
 \label{freens1}
 \ee
 { and $-{\blf k}'_{23}\equiv((M_{23}' - k_{23}^{'+}),-{\bf k}'_{23\perp})$.}
 
The  right-hand side of Eq. (\ref{lfwf3}) reflects: {(i) the normalization properties of 
$| {{\blf k}'_{1} }\rangle_{LF}$ and $| { \tilde{\bm \kappa}_1}\rangle_{LF}$; (ii) the expression for the intrinsic
wave function of the interacting pair (2,3); (iii)} the proper overall normalization factors. 

In Appendix \ref{appc} the correctness of the normalization factors in Eq. (\ref{lfwf3}) is checked.

 To obtain the last step in Eq.  (\ref{lfwf3}), one has to notice that the states
{ $| {{\blf k}_{1} \sigma_1}\rangle_{LF}$ and $| {\tilde{\bm \kappa}_1}\sigma_1\rangle_{LF}$
 are {immediately} {related to  the same LF state  $|\xi_1 ,{{\bm \kappa}_{1\perp}= 
  {\bf k}_{1\perp},\sigma_1 }\rangle$, } since   $\xi_1= 
 \kappa^+_1/{\cal M}_0(1,23)=k^{+}_1/M_0(1,2,3)$}. The two states  differ for their normalization, i.e. 
\be
 _{LF}\langle{\blf {k}}'_{1} | {\blf {k}}_{1}\rangle_{LF}= 
 {(2\pi)^3~2 k^{+}_1 }~ \delta^3({\blf k}'_1 -{\blf k}_1) 
 %= {(2\pi)^3~2 k^{+}_1 }~ \delta^3({\bf k}'_{1\perp} -{\bf k}_{1\perp})~\delta(\xi'_1 - \xi_1)(1-\xi_1)/%E_{23}
 \label{norm-k}
 \ee
 and 
 \be
 _{LF}\langle { \tilde{\bm \kappa}'_{1} }| {\tilde{\bm \kappa}_1}\rangle_{LF}= 
 {(2\pi)^3~2 \kappa^{+}_1 }~ \delta^3(\tilde{\bm \kappa}'_1 - \tilde{\bm \kappa}_1) \quad \quad .
 \label{norm-kappa}
 \ee
 In Eq.  (\ref{lfwf3}) {${ k}_1^{+(a)}$  is obtained 
 by transforming $\kappa^+_1$ from the frame where
   $P^+={\cal M}_0(1,23)$ to the frame where $P^+=M_0(1,2,3)$
   through  a longitudinal LF boost, while  
 ${\bf k}_{1\perp}^{(a)}$ remains unchanged, i.e.} one has 
 ${\bf k}_{1\perp}^{(a)} = \tilde{\bm \kappa}_{1\perp}$
  (see Eq. (\ref{nint})).
% , i.e. ${\bf \kappa}^+_1=\xi~ {\cal{M}}_0(1,23)$
 %with $\xi=k_1^{+(a)}/M_0(1,2,3)$, while $\kappa_{1\perp}={\bf k}_{1\perp}$.
  To determine $k_1^{+(a)}$ one can first evaluate ${\cal{M}}_0(1,23)$ from Eq. (\ref{clus2})
 \be
 {\cal{M}}_0(1,23)= {(\kappa_1^+)^2 + (m^2 + k^2_{1\perp}) \over 2 ~ \kappa_1^+ }
 ~+\left [ \left [(\kappa_1^+)^2 + (m^2 + k^2_{1\perp}) \over 2 ~ \kappa_1^+ \right ]^2 ~ +
 ~ M_S^2 ~ - ~ m^2 \right ]^{1/2} \quad .
 \label{CallM0}
 \ee
 Then one can obtain $\xi_1 $
 \be
 \xi_1 = {\kappa_1^+ \over {\cal{M}}_0(1,23)} \quad \quad ,
 \label{csi}
 \ee
 the  three-body system free mass $M_0(1,2,3)$
 \be
 M^2_0(1,2,3) = {m^2 + k^2_{1\perp} \over \xi_1} ~ + ~ {M'^2_{23} + k^2_{1\perp} \over 1 - \xi_1}
 \ee
 and
 \be
 k_1^{+(a)} ~ = ~ \xi_1 ~ M_0(1,2,3) \quad \quad .
 \label{kpiu}
 \ee

\subsubsection{Overlaps between {cluster states} and the bound-state of the three-particle system}
\label{overl}
{The
overlap between a state of the cluster $1+(2 3)$ and the bound state of the
three-particle system is the needed quantity for defining the LF spin-dependent
 spectral
 function. As a matter of fact, from  Eqs. (\ref{3bwf}) and (\ref{1e23}),}  one has  
\be   
_{LF}\langle   T_{23},
\tau_{23};\alpha,\epsilon_{23}; j_{23} j_{23z};\tau_1\sigma_1  
\tilde{\bm \kappa}_1; {\blf P}'| {\blf P}; j,j_z;
\epsilon^3_{int},\Pi; {1\over 2} T_z \rangle
=\nonu= 
2 P^+ (2\pi)^3~\delta^3({\blf P}'-{\blf P})~_{LF}\langle   T_{23},
\tau_{23};\alpha,\epsilon_{23}; j_{23} j_{23z};\tau_1\sigma_1  
\tilde{\bm \kappa}_1 |  j,j_z;
\epsilon^3_{int},\Pi; {1\over 2} T_z \rangle \quad .
\label{overlap}
\ee
As shown in Appendix \ref{appc2}, after inserting in the intrinsic part of the overlap (\ref{overlap})
(i) the completeness operator expressed through plane waves,
% for the intrinsic part, 
 i.e. (cf Eq. (\ref{compl3})) 
\be
\int {d{\blf  k}'_{23}\over  k^{\prime +}_{23}
(2\pi)^3 }| {\blf k}'_{23}\rangle~\langle{\blf k}'_{23}|
~\int {{{M}}'_0(1,2,3) ~d{\blf  k}'_1\over 2 k^{\prime +}_1 E'_{23}(2\pi)^3 }
~
 | {\blf k}'_{1}\rangle 
\langle{\blf k}'_1  | = {\bf I}  \quad ,
\label{complpw}
\ee
and  (ii) Eqs. (\ref{lfwf4})  and  (\ref{lfwf3}), one gets
{\be
 _{LF}\langle   T_{23},\tau_{23};\alpha,\epsilon_{23}; j_{23} j_{23z};\tau_1\sigma_1  
\tilde{\bm \kappa}_1 |  j,j_z;
\epsilon^3_{int},\Pi; {1\over 2} T_z \rangle
=
~ \sum_{\tau_2\tau_3}  ~
%\int { d{\blf  k}'_1 }~
 \int d {\bf k}_{23} ~ 
 ~\sum_{\sigma'_1}~
D^{{1 \over 2}} [{\cal R}_M ({\blf k}^{(a)}_1 )]_{\sigma_1\sigma'_1}\nonu \times ~
\sqrt{{(2\pi)^3 } ~2E({\bf k}^{(a)}_1)}~
\sqrt{\kappa_1^+ E_{23}\over k^{+(a)}_1 E_S}~
 \sum_{\sigma''_2,\sigma''_3}~\sum_{\sigma'_2,\sigma'_3}
~{\cal D}_{\sigma''_2,\sigma'_2}({\blf k}_{23},{\blf k}_{2})~
{\cal D}_{\sigma''_3,\sigma'_3}(-{\blf k}_{23},{\blf k}_{3})
\nonu \times ~ 
\langle   T_{23},
\tau_{23};\alpha,\epsilon_{23}; j_{23} j_{23z}| {\bf k}_{23}, \sigma''_2, \sigma''_3; \tau_2,\tau_3 \rangle
~
\langle \sigma'_3, \sigma'_2,\sigma'_1; \tau_3,\tau_2,\tau_1; {\bf k}_{23},{\bf k}^{(a)}_1|j,j_z;
\epsilon^3_{int},\Pi; {1\over 2} T_z \rangle ~ 
 \quad  ,
%= ~ \sum_{\tau_2\tau_3} \sum_{\sigma'_1}
%D^{{1 \over 2}} [{\cal R}_M(\tilde{\bm \kappa}_{1} )]_{\sigma_1\sigma'_1}~
%~\int d {\bf k}'_{23} ~ \sqrt{(2\pi)^3~2E({\bf k}_1^{(a)})}~\sqrt{\kappa_1^+ E'_{23}\over k^{+(a)}_1 %E_S}
%\nonu \times ~
%\sum_{\sigma'_2,\sigma'_3}\langle   T_{23},
%\tau_{23};\alpha,\epsilon_{23}; j_{23} j_{23z}| {\bf k}'_{23}, \sigma'_2, \sigma'_3; \tau_2,\tau_3 
%\rangle ~
%\langle \sigma'_3, \sigma'_2,\sigma'_1; \tau_3,\tau_2,\tau_1; {\bf k}'_{23},{\bf k}_1^{(a)}|j,j_z;
%\epsilon^3_{int},\Pi; {1\over 2} T_z \rangle \quad   ,
\label{ovrla}
\ee
where the unitary matrices ${\cal D}_{\sigma''_i,\sigma'_i}$ are defined by the equation
\be
{\cal D}_{\sigma''_i,\sigma'_i}
(\pm{\blf k}_{23},{\blf k}_{i})=
\sum_{\sigma_i}~
 D^{{1 \over 2}} [{\cal R}^\dagger_M{ (\pm{\blf k}_{23} )}]_{\sigma''_i\sigma_i}~
 D^{{1 \over 2}} [{\cal R}_M {({\blf k}_{i} )}]_{\sigma_i\sigma'_i}
 \quad \quad 
\ee}
with the $+$ sign corresponding to $i=2$ and the $-$ sign corresponding to $i=3$.

Then the overlap of Eq. (\ref{overlap}) can be evaluated by approximating  
$\langle   T_{23},\tau_{23};\alpha,\epsilon_{23}; j_{23} j_{23z}| 
{\bf k}_{23}, \sigma''_2, \sigma''_3; \tau_2,\tau_3 \rangle~$ and 
$~\langle \sigma'_3, \sigma'_2,\sigma'_1; \tau_3,\tau_2,\tau_1; {\bf k}_{23},{\bf k}_1|
j,j_z; \epsilon^3_{int},\Pi; {1\over 2} T_z \rangle~$  with the corresponding non relativistic quantities. It should be recalled that the
spins involved are canonical spins.}
%In view of the normalization (\ref{normhe3}), the above overlap could be
%approximated by a non relativistic one.
 
The normalization for the intrinsic LF overlap in Eq. (\ref{overlap})
follows immediately from the completeness relation (\ref{complns}) 
%or the definition (\ref{ovrl}) and Eq. (\ref{compl2b}), 
\be
\sum_{ T_{23}\tau_{23} \tau_1} \int {d\tilde{\bm \kappa}_{1}\over 2 \kappa^+_1 (2\pi)^3}
~\sumint {\lambda(t) ~ d{  t}
}~
%{2M_0~t^2 d{  t}\over E_{23} M_{23}(2\pi)^3 }
%\nonu \times ~
\sum_{\sigma_1} \sum_{j_{23}j_{23z} \alpha} \left |_{LF}\langle   T_{23},
\tau_{23};\alpha,\epsilon_{23}; j_{23} j_{23z};\tau_1\sigma_1,  \tilde{\bm \kappa}_1|
  j,j_z; \epsilon^3_{int},\Pi; {1\over 2} T_z \rangle \right|^2=
\nonu
%\sum_{\tau_1,\tau_2,\tau_3}\sum_{\sigma_1\sigma_2\sigma_3 }
%\int {d{\bf  \kappa}_{1}\over 2 E({\bf \kappa}_1) (2\pi)^3}2 
%E({\bf \kappa}_1) (2\pi)^3\int dj ~
%|\langle \tau_1,\tau_2,\tau_3,\sigma_1,\sigma_2,\sigma_3; {\bf \kappa}_1,j|j,j_z;
%\epsilon^3_{int},\Pi; {1\over 2} T_z \rangle|^2
= \left | | j,j_z; \epsilon^3_{int},\Pi; {1\over 2} T_z \rangle \right|^2 = ~ 1 \quad \quad .
\label{overlapNorm}
\ee
As shown in Appendix  \ref{appc3}, this normalization can be recovered using the explicit expression for the overlaps given in Eq. (\ref{ovrla}).

\section{The LF spin-dependent spectral function}  
\label{sf}
The non relativistic spin-dependent spectral function ${\bf\hat{P}}^{\tau}_{\cal{M}}(\vec{p},E)$
for a nucleus of mass number A is {a $2 \times 2$ matrix, whose 
elements are}
\be 
 P_{\sigma,\sigma',\cal{M}}^{\tau} ({\vec{p},E})=
 \sum\nolimits\limits_{{f}_{(A-1)}}
~\langle{\vec{p},\sigma \tau;\psi
}_{f_{(A-1)}} |{\psi }_{\cal{J}\cal{M}}\rangle~\langle{\psi
}_{\cal{J}\cal{M}}|{\psi }_{f_{(A-1)}};\vec{p},\sigma ' \tau \rangle ~  
\delta (E-{E}_{f_{(A-1)}}+{E}_{A}) \quad ,
\label{spec0}
\ee
where $|{\psi}_{\cal{J}\cal{M}}\rangle$ is the ground state of the nucleus with energy $E_A$ and polarized along
$\vec{S}$, $|{\psi }_{f_{(A-1)}}\rangle$  an eigenstate of the (A-1) nucleon
system with energy $E_{f_{(A-1)}}$, interacting with the same interaction of the nucleus,
$|\vec{p},\sigma \tau\rangle$  the plane wave for the nucleon $\tau = \pm 1/2$, with momentum 
$\vec{p}$ in the nucleus rest frame   and  spin
along the z-axis equal to $\sigma$ \cite{cps,cps1,kpsv}. 
The state $|{\psi}_{\cal{J}\cal{M}}\rangle$ polarized along $\vec{S}$ can be expressed through the states $|{\psi}_{{\cal{J}}{m}}\rangle_z$ polarized along the $z$ axis \cite{cps1,Varsa} as follows
\be
|{\psi}_{\cal{J}\cal{M}}\rangle_{\vec S} = \sum_m |{\psi}_{{\cal{J}}{m}}\rangle_z ~ 
D^{\cal J}_{m, \cal M}(\alpha, \beta, \gamma) \quad ,
\label{rot}
\ee
where $\alpha, \beta$ and $\gamma$ are the Euler angles describing the proper rotation from the $z$-axis 
to the polarization vector $\vec S$. {Let us recall that the rotations involved act on the three-nucleon bound
system as a whole,
and therefore they are interaction-free.}

In a more compact form, for ${\cal{J}}=1/2$, the  $2\times 2$  matrix
${\bf\hat{P}}^{\tau}_{\cal{M}}(\vec{p},E)$ is given by  
\be
{\bf\hat{P}}^{\tau}_{\cal{M}}(\vec{p},E)={1\over
2}\left[B_{0,{\cal{M}}}^{\tau}(|\vec{p}|,E)~+~
\vec{\sigma} \cdot \vec{f}^{\tau}_{\cal
{M}}(\vec{p},E)\right] \quad ,
\label{spec1}
\ee
where the function $B_{0,\cal{M}}^{\tau}(|\vec{p}|,E)$ is the trace of 
${\bf\hat{P}}^{\tau}_{\cal{M}}(\vec{p},E)$ and yields the
usual unpolarized spectral function $P^{\tau}\!\left( |\vec{p}| , E\right)$. 
It should be noticed that the matrix
${\bf\hat{P}}^{\tau}_{\cal{M}}(\vec{p},E)$ and the pseudovector
$\vec{f}^{\tau}_{\cal{M}}(\vec{p},E)$ depend on the direction of the polarization
vector $\vec{S}$. 
Since   $\vec{f}^{\tau}_{\cal{M}}(\vec{p},E)$ is a 
pseudovector, it is a linear combination of the pseudovectors at our disposal,
viz. $\vec{S}$ and $\hat{p}~(\hat{p} \cdot \vec{S})$, and therefore it can
be put in the following form, where any angular dependence is explicitely given,

\be
\vec{f}^{\tau}_{\cal{M}}(\vec{p},E)~=~
\vec{S}~B_{1,\cal{M}}^{\tau}(|\vec{p}|,E)~+~\hat{p}~(\hat{p}
\cdot \vec{S})~ B_{2,\cal{M}}^{\tau}(|\vec{p}|,E) \quad .
\label{spec2}
\ee

{Let us focus on the $A=3$ case.} To obtain a Poincar\'e covariant definition of the spin-dependent spectral function 
for a three-particle system within the LF dynamics, one replaces the non relativistic overlaps 
$\langle{\vec{p},\sigma \tau;\psi}_{f_{(A-1)}} |{\psi }_{\cal{J}\cal{M}}\rangle$,
which define the non relativistic spectral function,
with their LF counterparts 
$_{LF}\langle \tau_{S},T_{S};\alpha,\epsilon ;J_{z}J;\tau\sigma,\tilde{\bm \kappa}|\Psi_{0}; S, T_z\rangle$,
dependent upon the energy $\epsilon$ of the two-body system and upon the intrinsic momentum, 
$\tilde{\bm \kappa}$, of the third particle   in the intrinsic reference frame of the cluster $1+(23)$ (cf
 Sec. \ref{overl}).
The LF overlaps $_{LF}\langle \tau_{S},T_{S};\alpha,\epsilon ;J_{z}J;\tau\sigma,\tilde{\bm \kappa}|\Psi_{0}; S, T_z\rangle$
can be easily obtained  from the overlaps of Eq.  (\ref{ovrla}), writing through Eq. (\ref{rot}) the ground  state 
$|\Psi_{0};S, T_z \rangle$ of the three-body system, polarized along $\vec S$, in terms of the states
$|j,j_z; \epsilon^3_{int},\Pi; {1\over 2} T_z \rangle$, polarized along the $z$ axis.

 Then, within the LFHD one can define  the spin-dependent nucleon spectral function for 
 the three-nucleon system ($^3$He or $^3$H) 
 in the bound state $|\Psi_{0};S, T_z \rangle$,
 as follows
\be
 {\cal P}^{\tau}_{\sigma'\sigma}(\kappa^+,{\bm \kappa}_\perp,\kappa^-,S)
= 
%\left|{\partial \kappa^+\over \partial \xi}\right|
~\sumint  d\epsilon~
\rho(\epsilon) 
~\delta\left( \kappa^- -M_3+{M^2_S +|{\bm \kappa}_\perp|^2 \over (1-\xi)M_3}
\right) 
\nonu
\times 
\sum_{J J_{z}\alpha}\sum_{T_{S}\tau_{S} } ~
_{LF}\langle  \tau_{S},T_{S} , 
\alpha,\epsilon; J J_{z}; \tau\sigma',\tilde{\bm \kappa}|\Psi_{0}; S,T_z
\rangle
  ~\langle S,T_z;
\Psi_0|\tilde{\bm \kappa},\sigma\tau; J J_{z}; 
\epsilon, \alpha, T_{S}, \tau_{S}\rangle_{LF} \nonu 
= {1 \over  \left | { \partial \kappa^- \over \partial \epsilon} \right | } ~ \rho(\epsilon) 
\sum_{J J_{z}\alpha}\sum_{T_{S}\tau_{S} } ~
_{LF}\langle  \tau_{S},T_{S} ; 
\alpha,\epsilon ;J J_{z}; \tau\sigma',\tilde{\bm \kappa}|\Psi_{0}; S, T_z\rangle
~ \langle S,T_z;
\Psi_0|\tilde{\bm \kappa},\sigma\tau; J J_{z}; 
\epsilon, \alpha; T_{S}, \tau_{S}\rangle_{LF} \nonu
\hspace{-0.4cm}  
=  \left | { \partial \epsilon \over \partial \kappa^-} \right | ~
{\cal P}^{\tau}_{\sigma'\sigma}(\tilde{\bm \kappa},\epsilon,S) \quad ,
\label{LFspf}
\ee
where 
\be
\epsilon ~ = ~ { (M_3 - \kappa^-) (1-\xi) M_3  - |{\bm  \kappa}_{\perp}|^2 \over  4 ~ m} - m ~
% = ~ t^2/m
\label{epsi3}
\ee
is the intrinsic energy of the fully interacting two-nucleon eigenstate,
 $\rho(\epsilon)$ the density of the two-body states  
($\rho(\epsilon)  =  tm/2$ for the two-body continuum states and
$\rho(\epsilon)  = 1$ for the deuteron  bound state),
 $M_3$  the nucleus mass, {$\xi=\kappa^+/{\cal M}_0(1,23)$ (cf Eqs. (\ref{nint}) and 
 (\ref{CallM0}))} and 
\be
\left | { \partial \epsilon \over \partial \kappa^-} \right | = {(1-\xi) M_3  \over 4m} \quad .
\label{devke}
\ee 
Let us notice that the variable $\kappa^-$ is {the $-$ component of the momentum of an off mass shell nucleon}, as it is clear from the $\delta$ function in Eq. (\ref{LFspf}).
In Eq.  (\ref{LFspf}) $\tau= \pm 1 /2$,
 $J, ~ J_{z}$ is the spin, $T_{S}, ~ \tau_{S}$ the isospin, 
% $\epsilon=t^2/m$ , 
 %and
 $\alpha$ the set of quantum numbers needed to completely specify the   two-body
%  $(N-1)$-nucleon 
  eigenstate, and $M_S^2 = 4(m^2 + m \epsilon)$. 
  
The overlap 
$_{LF}\langle \tau_{S},T_{S};\alpha,\epsilon ;J_{z}J;\tau\sigma,\tilde{{\bm \kappa}}|\Psi_{0}; S, T_z\rangle$ is the one defined by Eqs. (\ref{rot}) and (\ref{ovrla}). 
%Therefore, i
In the special case where $\vec S$ is along the $z$-axis, 
%using the explicit expression for this overlap given in Eq. (\ref{ovrla}) in terms of canonical two- %and three-body wave functions, 
one obtains
\be
{\cal P}^{\tau}_{\sigma'\sigma}(\tilde{\bm \kappa},\epsilon,S) =
\rho(\epsilon) \nonu
\times ~
\sum_{J J_{z}\alpha}\sum_{T_{S}\tau_{S} } ~
_{LF}\langle  \tau_{S},T_{S} ; 
\alpha,\epsilon ;J J_{z}; \tau\sigma',\tilde{\bm \kappa}|
j,j_z;
\epsilon^3_{int},\Pi; {1\over 2} T_z\rangle
~ \langle{1\over 2} T_z; \Pi,\epsilon^3_{int}; j,j_z; 
|\tilde{\bm \kappa},\sigma\tau; J J_{z}; 
\epsilon, \alpha; T_{S}, \tau_{S}\rangle_{LF} 
\nonu
%%
%~ \sum_{\sigma'_1}
%D^{{1 \over 2}} [{\cal R}_M(\tilde{\bm \kappa}_{1} )]_{\sigma' \sigma'_1}~
% \sum_{\sigma''_1}
%D^{{1 \over 2} *} [{\cal R}_M(\tilde{\bm \kappa}_{1} )]_{\sigma \sigma''_1}~
%\nonu \times ~
%\sum_{J J_{z}\alpha}\sum_{T_{S}\tau_{S} }  ~
%\sum_{\tau'_2\tau'_3}
% ~\int d {\bf k}'_{23} ~ \sqrt {E({\bf k}_1^{'(a)}) ~ E'_{23}\over k^{'+(a)}_1}
%~\sum_{\sigma'_2,\sigma'_3}\langle   T_{S},
%\tau_{S};\alpha,\epsilon; J J_{z}| {\bf k}'_{23}, \sigma'_2, \sigma'_3; \tau'_2,\tau'_3 \rangle
%~\nonu \times ~
%\langle \sigma'_3, \sigma'_2,\sigma'_1; \tau'_3,\tau'_2,\tau; {\bf k}'_{23},{\bf k}_1^{'(a)})|j,j_z;
%\epsilon^3_{int},\Pi; {1\over 2} T_z \rangle ~  \nonu \times ~
%\sum_{\tau''_2\tau''_3}
%~\int d {\bf k}''_{23} ~ \sqrt {E({\bf k}_1^{''(a)}) ~ E''_{23}\over k^{''+(a)}_1}
%~\sum_{\sigma''_2,\sigma''_3}\langle   T_{S},
%\tau_{S};\alpha,\epsilon; J J_{z}| {\bf k}''_{23}, \sigma''_2, \sigma''_3; \tau''_2,\tau''_3 \rangle ^*
%~\nonu \times ~
%\langle \sigma''_3, \sigma''_2,\sigma''_1; \tau''_3,\tau''_2,\tau; {\bf k}''_{23},{\bf k}_1^{''(a)})|j,j_z;
%\epsilon^3_{int},\Pi; {1\over 2} T_z \rangle ^*
\label{SFex}
\ee
and the LF spectral function can be evaluated through the explicit expression (\ref{ovrla}) for the overlap $_{LF}\langle  \tau_{S},T_{S} ; \alpha,\epsilon ;J J_{z}; \tau\sigma,\tilde{\bm \kappa}|
j,j_z;\epsilon^3_{int},\Pi; {1\over 2} T_z\rangle$
in terms of canonical two- and three-body wave functions. In turn, these wave functions can be replaced by the non relativistic ones. We emphasize once more that the two- and three-body non relativistic wave functions  have all the needed properties with  respect 
to rotations and translations  of the corresponding canonical wave functions.

According to the completeness relation (\ref{complns}), the normalization of the spectral function reads  
(see also Eq.  (\ref{overlapNorm}) 
and Appendix
\ref{appc})
\be
 \sumint {d\epsilon 
 }
 \int {d {\bm \kappa} \over 2E({\bf \kappa}) 
  (2\pi)^3}~ 
\sum _{\tau} Tr {\cal P}^{\tau}(\tilde{{\bm \kappa}},\epsilon,S)~=~1 \quad \quad .
\label{normFSLF}
\ee  
However, in  applications one can normalize  the spectral function ${\cal P}^{\tau}(\tilde{{\bm \kappa}},\epsilon,S)$ 
 for each isospin channel, i.e.,
\be
 \sumint {d\epsilon
  }
  \int {d {\bm \kappa} \over 2E({\bf \kappa}) 
  (2\pi)^3}~ 
Tr {\cal P}^{\tau}(\tilde{{\bm \kappa}},\epsilon,S)~=~1 \quad \quad .
\label{normFSLF1}
\ee  

%e INUTILE CON LA NUOVA VERSIONE: The dependence of the spectral function from the %nucleon momentum $\bf p_1$ in the laboratory frame can be easily recovered. Indeed from the %values of $\bm  \kappa$ and $\epsilon$ one can define ${\cal{M}}_0(1,23)$ and $\xi_1$ through %Eqs. (\ref{CallM0}) and (\ref{csi}) and then 
%$p_1^+$, ${\bf p}_{1 \perp }$, and $p_1^-$ through Eqs. (\ref{cons1}) 
%and (\ref{meno}).}

%Using the momentum $\vec{p}$ and the energy $E$, 
%one can define $\xi = p^+ /P^+$, ${\cal{M}}_0(1,23)$ through Eq. (\ref{clus1}) 
%and $\tilde{\bm \kappa}_1$ through Eq. (\ref{nint}).

As it occurs for the non relativistic spectral function (see Eqs. (\ref{spec1}) and (\ref{spec2})), the LF nucleon spin-dependent spectral function can be expressed by means of three scalar functions,
${\cal {B}}_{0,S}^{\tau}(|{\bm \kappa}|,\epsilon)$, ${\cal {B}}_{1,S}^{\tau}(|{\bm \kappa}|,\epsilon)$ and ${\cal {B}}_{2,S}^{\tau}(|{\bm \kappa}|,\epsilon)$,  
%as follows
\be
 \hspace{-0.9cm}{\cal P}^{\tau}_{\sigma'\sigma}({\tilde{\bm \kappa}},\epsilon,S)=
%(2\pi)^3  {E({\bf \kappa}) }[{\cal{N}}(\epsilon)]^2\sum_{\sigma_1 \sigma'_1} 
%D^{{1 \over 2}} [{\cal R}_M^\dagger ({\blf \kappa})]_{\sigma'\sigma'_1}
{1 \over 2} ~ \left[{\cal {B}}_{0,S}^{\tau}(|{\bm \kappa}|,\epsilon)~+~
{\bm \sigma} \cdot {\bf f}^{\tau}_{S}({\bm\kappa},\epsilon) \right]_{\sigma'\sigma} \quad ,
%\nonu \times ~
%D^{{1 \over 2}} [{\cal R}_M ({\blf \kappa})]_{\sigma_1\sigma}
\label{speclf1}
\ee
where
\be
{\bf f}^{\tau}_{S}({\bm \kappa},\epsilon) ~=~
{\bf S}~{\cal {B}}_{1,S}^{\tau}(|{\bm \kappa}|,\epsilon)~+~\hat{\kappa}~(\hat{\kappa}
\cdot {\bf S})~ {\cal {B}}_{2,S}^{\tau}(|{\bm \kappa}|,\epsilon) \quad.
\ee
The function ${\cal{B}}_{0,S}^{\tau}(|{\bm \kappa}|,\epsilon)$ is the trace of 
${\cal P}^{\tau}_{\sigma'\sigma}({\tilde{\bm \kappa}},\epsilon,S)$ and yields the unpolarized spectral function. 

\subsection{The LF nucleon {momentum distributions} and momentum sum rule} 
 Within the LFHD, one can define  the LF spin-independent nucleon momentum distribution, averaged on the spin directions,
 through the spectral function ${\cal P}^{\tau}_{\sigma'\sigma}(\tilde{\bm \kappa},\epsilon,S)$ as follows
\be
n^\tau(\xi,{\bf k}_{\perp}) ~ = ~ \sumint {d\epsilon ~
 }
 {1 \over 2 \kappa^+~ (2\pi)^3}~ 
 ~ {\partial \kappa^+ \over \partial \xi} 
 ~
 %\sum _{\tau} 
 Tr {\cal P}^{\tau}(\tilde{\bm \kappa},\epsilon,S) =
\sumint {d\epsilon ~
} 
{1 \over 2 ~ (2\pi)^3}~
{ E_S \over (1- \xi)  ~ \kappa^+} ~ 
 \nonu \times ~
%{t m \over 2 }
\rho (\epsilon)
 ~ \sum _{\sigma}  
\sum_{J J_{z}\alpha}\sum_{T_{S}\tau_{S} } ~
_{LF}\langle  \tau_{S},T_{S} ; 
\alpha,\epsilon ;J_{z}J; \tau\sigma,\tilde{\bm \kappa}|\Psi_{0}; S T_z\rangle
~ \langle S,T_z;
\Psi_0|\tilde{\bm \kappa},\sigma\tau; J J_{z}; 
\epsilon, \alpha; T_{S}, \tau_{S}\rangle_{LF}
\quad ,
\label{momdisLF}
\ee  
where Eq. (\ref{devkappa}) has been used. 
%the equality (see Eqs. (\ref{devkappa}) and  (\ref{devkappa1}))
%\be
%{\partial \kappa_z \over \partial \xi} ~ = ~ { E_S ~ E(\bf \kappa) \over (1- \xi)  ~ \kappa^+}
%\ee
%and
From the completeness relation (\ref{complns}), one gets immediately the normalization of the  nucleon momentum distribution
\be
\int d \xi ~ \int d{\bf k}_{\perp} ~ n^\tau(\xi,{\bf k}_{\perp}) ~ =~1 \quad .
\label{momnorm}
\ee  
An explicit expression for the spin-averaged momentum distribution can be obtained inserting 
in Eq.  (\ref{momdisLF}) the {LF 
spectral function as written} in Eq. (\ref{SFex}) {and in turn the expression for the overlaps 
given in Eq. (\ref{ovrla}).}
%\be
%n^\tau(\xi,{\bf k}_{\perp}) ~ =  ~\sum _{\sigma} ~
%\sumint {d\epsilon \over  (1- \xi)  } ~
%\rho(\epsilon)  
%~ \sum_{\sigma'_1}
%D^{{1 \over 2}} [{\cal R}_M(\tilde{\bm \kappa}_{1} )]_{\sigma \sigma'_1}~
 %\sum_{\sigma''_1}
%D^{{1 \over 2} *} [{\cal R}_M(\tilde{\bm \kappa}_{1} )]_{\sigma \sigma''_1}~
%\nonu \times ~
%\rho({\bf k}_1,\epsilon_{23})
%\sum_{J J_{z}\alpha}\sum_{T_{S}\tau_{S} }  ~
%\sum_{\tau'_2\tau'_3}
 %~\int d {\bf k}'_{23} ~ \sqrt {E({\bf k}_1^{'(a)}) ~ E'_{23}\over k^{'+(a)}_1}
%~\sum_{\sigma'_2,\sigma'_3}\langle   T_{S},
%\tau_{S};\alpha,\epsilon; J J_{z}| {\bf k}'_{23}, \sigma'_2, \sigma'_3; \tau'_2,\tau'_3 \rangle
%~\nonu \times ~
%\langle \sigma'_3, \sigma'_2,\sigma'_1; \tau'_3,\tau'_2,\tau; {\bf k}'_{23},{\bf k}_1^{'(a)}|j,j_z;
%\epsilon^3_{int},\Pi; {1\over 2} T_z \rangle ~  \nonu \times ~
%\sum_{\tau''_2\tau''_3}
%\rho({\bf k}_1,\epsilon_{23})
%~\int d {\bf k}''_{23} ~ \sqrt {E({\bf k}_1^{''(a)}) ~ E''_{23}\over k^{''+(a)}_1}
%~\sum_{\sigma''_2,\sigma''_3}\langle   T_{S},
%\tau_{S};\alpha,\epsilon; J J_{z}| {\bf k}''_{23}, \sigma''_2, \sigma''_3; \tau''_2,\tau''_3 \rangle ^*
%~\nonu \times ~
%\langle \sigma''_3, \sigma''_2,\sigma''_1; \tau''_3,\tau''_2,\tau; {\bf k}''_{23},{\bf k}_1^{''(a)}|j,j_z;
%\epsilon^3_{int},\Pi; {1\over 2} T_z \rangle ^*
%\label{mdex}
%\ee
Then, using again  the two-body completeness of Eq.  (\ref{nrcompl1}) and the unitarity of the 
{${\cal D}$ and} $D^{1/2}$ matrices, one obtains 
\be
\hspace{-.9cm} n^\tau(\xi,{\bf k}_{\perp})  
 %~ {1  \over  (1- \xi)  } ~ \sum _{\sigma} ~ 
%\sum_{\tau'_2\tau'_3} ~\sum_{\sigma'_2,\sigma'_3}
 %~\int d {\bf k}'_{23} ~  {E({\bf k}_1^{'(a)}) ~ E'_{23}\over k^{'+(a)}_1}
%~\nonu \times ~
%\langle \sigma'_3, \sigma'_2,\sigma; \tau'_3,\tau'_2,\tau; {\bf k}'_{23},{\bf k}_1^{'(a)}|j,j_z;
%\epsilon^3_{int},\Pi; {1\over 2} T_z \rangle ~ 
% \nonu \times ~
%\langle \sigma'_3, \sigma'_2,\sigma; \tau'_3,\tau'_2,\tau; {\bf k}'_{23},{\bf k}_1^{'(a)}|j,j_z;
%\epsilon^3_{int},\Pi; {1\over 2} T_z \rangle ^* ~ = 
%\nonu
 =   {1  \over  1- \xi  } ~ \sum _{\sigma}  
%\rho({\bf k}_1,\epsilon_{23})
\sum_{\tau'_2\tau'_3} \sum_{\sigma'_2,\sigma'_3}
 \int d {\bf k}_{23} ~  {E({\bf k}_1) ~ E_{23}\over k^+_1}
~
%\nonu \times ~
\left | \langle \sigma'_3, \sigma'_2,\sigma; \tau'_3,\tau'_2,\tau; {\bf k}_{23},{\bf k}_1|j,j_z;
\epsilon^3_{int},\Pi; {1\over 2} T_z \rangle \right |^2 \quad ,
%\langle \sigma'_3, \sigma'_2,\sigma; \tau'_3,\tau'_2,\tau; j,{\bf k}_1|j,j_z;
%\epsilon^3_{int},\Pi; {1\over 2} T_z \rangle ^*
\label{mdex1}
\ee
where ${\bf k}_{1\perp} = {\bf k}_{\perp}$ and $k^+_1 = \xi ~ M_0(1,2,3)$ (see Eq. (\ref{kpiu})).
Combining Eqs.  (\ref{deracompl3}) and  (\ref{deracompl4}), the normalization of the LF nucleon momentum distribution (\ref{momnorm}) can be rewritten as follows 
\be
\int d \xi ~ \int d{\bf k}_{\perp} ~ n^\tau (\xi,{\bf k}_{\perp}) ~ =
~  \int d{\bf k}_{\perp} \sum _{\sigma} ~ 
\sum_{\tau_2\tau_3} ~\sum_{\sigma_2,\sigma_3}
 ~\int d {\bf k}_{23} ~    \int {\partial  \xi \over \partial {k_z}} ~ d k_z ~ 
 {\partial k_{z} \over \partial k^+} ~{E_{23}  \over  (1- \xi)  } ~
% { E({\bf k}) ~ \over k^+}
~\nonu \times ~
\left | \langle \sigma_3, \sigma_2,\sigma; \tau_3,\tau_2,\tau; {\bf k}_{23},{\bf k}|j,j_z;
\epsilon^3_{int},\Pi; {1\over 2} T_z \rangle \right |^2~=
%\langle \sigma_3, \sigma_2,\sigma; \tau_3,\tau_2,\tau; j,{\bf k}|j,j_z;
%\epsilon^3_{int},\Pi; {1\over 2} T_z \rangle ~ \langle \sigma_3, \sigma_2,\sigma; \tau_3,\tau_2,\tau; j,{\bf k}|j,j_z;
%\epsilon^3_{int},\Pi; {1\over 2} T_z \rangle ^* ~ =
 \nonu
=
~  \int d{\bf k}_{\perp} \sum _{\sigma} ~ 
\sum_{\tau_2\tau_3} ~\sum_{\sigma_2,\sigma_3}
 ~\int d {\bf k}_{23} ~    \int  ~ d k_z ~ 
\left | \langle \sigma_3, \sigma_2,\sigma; \tau_3,\tau_2,\tau; {\bf k}_{23},{\bf k}|j,j_z;
\epsilon^3_{int},\Pi; {1\over 2} T_z \rangle \right |^2~ = 
 \nonu
= ~  \int d{\bf k}_{\perp}  ~  \int  ~ d k_z ~ f^\tau(k_z,{\bf k}_{\perp})~ = ~1 \quad ,
\label{momnorm1}
\ee  
where $f^\tau(k_z,{\bf k}_{\perp})$ is  the {instant form}
 momentum distribution in terms of the intrinsic nucleon momentum
  {${\bf k} = {\bf k}_1$}, defined by Eqs.  (\ref{cons1}), and (\ref{kzz}) 
of Sec. \ref{lfk},
\be
f^\tau(k_z,{\bf k}_{\perp}) ~ = \sum _{\sigma} ~ 
\sum_{\tau_2\tau_3} ~\sum_{\sigma_2,\sigma_3}
 ~\int d {\bf k}_{23} ~   
\left | \langle \sigma_3, \sigma_2,\sigma; \tau_3,\tau_2,\tau; {\bf k}_{23},{\bf k}|j,j_z;
\epsilon^3_{int},\Pi; {1\over 2} T_z \rangle \right |^2 \quad .
\label{mdex2}
\ee

Let us show that the momentum sum rule 
\be
\int \xi ~ d \xi ~ \int d{\bf k}_{\perp} ~ n^\tau(\xi,{\bf k}_{\perp}) ~ =~{ 1 \over 3}
\label{momsumrule}
\ee
is satisfied by the LF momentum distribution $n^\tau(\xi,{\bf k}_{\perp})$.
Indeed, because of the symmetry of the three-body bound state, one has 
\be
\int \xi ~ d \xi ~ \int d{\bf k}_{\perp} ~ n^\tau(\xi,{\bf k}_{\perp}) ~ = \nonu
=  ~
\sum_{\tau_2\tau_3} ~\sum_{\sigma_1\sigma_2,\sigma_3}
   \int   ~  d{\bf k}_1  \int d {\bf k}_{23} ~ {k^+_1 \over M_0(1,2,3)} ~
\left | \langle \sigma_3, \sigma_2,\sigma_1; \tau_3,\tau_2,\tau; {\bf k}_{23},{\bf k}_1|j,j_z;
\epsilon^3_{int},\Pi; {1\over 2} T_z \rangle \right |^2  = \nonu
=  ~
\sum_{\tau_2\tau_3} ~\sum_{\sigma_1\sigma_2,\sigma_3}
   \int   ~  d{\bf k}_2  \int d {\bf k}_{31} ~ {k^+_2 \over M_0(1,2,3)} ~
\left | \langle \sigma_3, \sigma_2,\sigma_1; \tau_3,\tau_2,\tau; {\bf k}_{31},{\bf k}_2|j,j_z;
\epsilon^3_{int},\Pi; {1\over 2} T_z \rangle \right |^2  = \nonu
=  ~
\sum_{\tau_2\tau_3} ~\sum_{\sigma_1\sigma_2,\sigma_3}
   \int   ~  d{\bf k}_3  \int d {\bf k}_{12} ~ {k^+_3 \over M_0(1,2,3)} ~
\left | \langle \sigma_3, \sigma_2,\sigma_1; \tau_3,\tau_2,\tau; {\bf k}_{12},{\bf k}_3|j,j_z;
\epsilon^3_{int},\Pi; {1\over 2} T_z \rangle \right |^2  = \nonu
={ 1 \over 3}  \sum_{\tau_2\tau_3} \sum_{\sigma_1\sigma_2,\sigma_3}
   \int   ~  d{\bf k}_1  \int d {\bf k}_{23} ~ {(k^+_1 + k^+_2 + k^+_3)\over M_0(1,2,3)} ~
\left | \langle \sigma_3, \sigma_2,\sigma_1; \tau_3,\tau_2,\tau; {\bf k}_{23},{\bf k}_1|j,j_z;
\epsilon^3_{int},\Pi; {1\over 2} T_z \rangle \right |^2  = \nonu
=~{ 1 \over 3}  \quad \quad  ,
\label{momsumrule1}
\ee
since (see Eqs.  (\ref{compl4}), and (\ref{lfwf4}) )
\be
\hspace{-1.cm} \left[{\partial ({\bf k}_1,{\bf k}_{23} ) \over \partial ({\bf k}_2,{\bf k}_{31} )}\right]=
{M_{23}~ E_1 ~ E_{23}\over M_{31}~ E_2 ~ E_{31}}~   \quad , \quad
\left[{\partial ({\bf k}_1,{\bf k}_{23} ) \over \partial ({\bf k}_3,{\bf k}_{12} )}\right]=
{M_{23}~ E_1 ~ E_{23}\over M_{12}~ E_3 ~ E_{12}}~ \quad ,  
\ee
\be
\hspace{-1.cm}\sqrt{E_1 ~ E_{23} ~ M_{23}} ~ | {\bf k}_1, {\bf k}_{23} \rangle = 
\sqrt{E_2 ~ E_{31} ~ M_{31}} ~  | {\bf k}_2, {\bf k}_{31}\rangle =
\sqrt{E_3 ~ E_{12} ~ M_{12}}  ~ | {\bf k}_3, {\bf k}_{12}\rangle \quad ,
\label{momsumrule2}
\ee
and $~k_1^+ ~+ ~  k_2^+ ~+ ~ k_3^+ ~= M_0(1,2,3)$. 
The momentum sum rule, Eq. (\ref{momsumrule}) has also been successfully checked 
calculating numerically Eq. (\ref{momsumrule1}) in an actual case  using the three-body wave-function of Ref. \cite{pisa} with the nuclear interaction of Ref. 
\cite{AV18}. {In the case of the proton {(with
accuracy produced by the normalization of the non relativistic wave function)} we obtain 0.9989 for 
the normalization and  0.3324 for the sum rule, while for the neutron we have    0.9981  and  0.3336, respectively
(see also Ref. \cite{Rinaldi}).}

Within the BT framework one can obtain LF momentum distributions dependent upon the spin directions, $n^\tau_{\sigma ' \sigma }(\xi,{\bf k}_{\perp};\vec S) $,  for any direction of the polarization vector $\vec S$ of the three-body system,
% and for any direction of the polarization $\vec n$ of the spin of the constituent
 using Eq. (\ref{rot}) and the expression for the LF spin-dependent spectral function given by Eq. (\ref{SFex}) 

\be
\hspace{-.8cm} n^\tau_{\sigma ' \sigma }(\xi,{\bf k}_{\perp};\vec S) ~ =  
{1 \over  (1- \xi)  } ~
 \sum_{\tau_2\tau_3}  ~
\int d {\bf k}_{23} ~ 
~\sum_{\sigma'_1}~
D^{{1 \over 2}} [{\cal R}_M ({\blf k}^{(a)}_1 )]_{\sigma'\sigma'_1}~
{E({\bf k}^{(a)}_1)}~ {E_{23}\over k^{+(a)}_1}
\nonu
 {\hspace{-.8cm} \times
\sum_{\sigma'_2,\sigma'_3}~ 
 } 
 \sum_m  ~ 
D^{j}_{m, {\cal M}}(\alpha, \beta, \gamma) ~
\langle \sigma'_3, \sigma'_2,\sigma'_1; \tau_3,\tau_2,\tau; {\bf k}_{23},{\bf k}^{(a)}_1|j,j_z = m;
\epsilon^3_{int},\Pi; {1\over 2} T_z \rangle ~
 \nonu 
 \hspace{-.8cm} \times ~
   \sum_{{\bar \sigma'_1}}
D^{{1 \over 2}*} [{\cal R}_M ({\blf k}^{(a)}_1 )]_{\sigma{\bar \sigma}'_1}~  
 \sum_{m'}   [D^{j}_{m', {\cal M}}(\alpha, \beta, \gamma)]^*\langle 
 {\sigma}'_3, {\sigma}'_2,{\bar \sigma}'_1; {\tau}_3,{\tau}_2,{\tau}; 
{\bf k}_{23},{\bf k}^{(a)}_1|j,j_z = m';
\epsilon^3_{int},\Pi; {1\over 2} T_z \rangle^*    ~~ .
\label{mdexspin}
\ee
{We remind that } $\alpha, \beta$ and $\gamma$ are the Euler angles describing the rotation from the $z$-axis 
to the polarization vector $\vec S$.} 
 In Eq. (\ref{mdexspin}) the explicit expression (\ref{ovrla}) for the overlaps is used, as well as the two-body completeness and  once again the unitarity of the {${\cal D}$ and} $D^{1/2}$ matrices.}

      \section{Conclusions and perspectives}
\label{concl}      
 In this paper, within the BT approach for the Poincar\'e generators, a LF 
 spin-dependent spectral  function and  LF  spin-dependent momentum  
 distributions have been defined starting from the LF wave function for a 
 three-body system, having in mind the
  $^3$He and the $^3$H nuclei. {The spectral function is defined through the overlaps between the ground state wave function of the three-body system  and the tensor product of a plane wave for one of the nucleons {in the intrinsic reference frame of the cluster (1,23)} and the state which describes the intrinsic motion of 
   the fully interacting {two-nucleon spectator} subsystem.}
In the present approach the packing operators, 
  needed to implement the macrocausality, are  not considered in the 
  description of the ground state of the three-body system, but the
   macrocausality is fully considered in  the mentioned tensor 
   product.  
   
 A generalization to $A$-nucleon nuclei is straightforward: one has only to 
 generalize the definition of the intrinsic momentum $\bf \kappa$ as 
 the momentum of one of the nucleons in the intrinsic reference frame of the cluster composed by this free nucleon and by the fully interacting system of the remaining $A-1$ nucleons. Then the LF spin-dependent spectral  function for the $A$-nucleon nucleus is
 \be
 \hspace{-0.5cm}{{{\cal {P}}^{\tau}_{\sigma'\sigma}(\kappa^+,{\bm \kappa}_\perp,\kappa^-,S,A)}
= 
%\left|{\partial \kappa^+\over \partial \xi}\right|
~\sumint  d\epsilon_{A-1}~
{\rho(\epsilon_{A-1})_{A-1} }
~\delta\left( \kappa^- -M_A+{M^2_{A-1} +|{\bm \kappa}_\perp|^2 \over (1-\xi)M_A}
\right)} ~ \times 
\nonu
 \hspace{-0.5cm}{{\sum_{J J_{z}\alpha}\sum_{T_{{A-1} }\tau_{{A-1} } } ~\hspace{-.4cm}
_{LF}\langle  \tau_{{A-1} },T_{{A-1} } , 
\alpha,\epsilon_{A-1}; J J_{z}; \tau\sigma',\tilde{\bm \kappa}|A, \Psi_{0}; S,T_z
\rangle
  ~\langle S,T_z;
\Psi_0, A|\tilde{\bm \kappa},\sigma\tau; J J_{z}; 
\epsilon_{A-1}, \alpha, T_{{A-1} }, \tau_{{A-1} }\rangle_{LF}}}
\nonu
\ee
where  $|A, \Psi_{0}; S,T_z \rangle$ is the ground-state of the $A$-nucleon nucleus, while
$M_{A-1}$ and $\epsilon_{A-1}$ are the mass and the intrinsic energy, 
{$\rho (\epsilon_{A-1})_{A-1}$} is the density, $J, J_z$  the spin, 
$T_{{A-1} }, \tau_{{A-1} }$ the isospin
 of the $(A-1)$-nucleon system and $\alpha$ the set of quantum numbers needed to fully specify %the $(A-1)$-nucleons 
 this system.
 
{Notably} within 
the LF Hamiltonian dynamics, both normalization and momentum sum rule 
can be exactly satisfied at the same time.
With respect to previous attempts to describe DIS processes 
off $^3$He in a LF framework (see, e.g., the one in Ref. \cite{OSC}), 
in our approach for the spin-dependent spectral function a 
special care is devoted to the definition of the intrinsic LF 
variables of the problem, as well as to the spin degrees of freedom 
through the Melosh rotations.
 
Our approach allows one to embed in a Poincar\'e covariant framework the large amount of knowledge on the nuclear 
interaction obtained from the non relativistic description of nuclei, since we adopt the LF version of the relativistic
 Hamiltonian dynamics with a fixed number of on-mass-shell constituents. The LF form of RHD has a sub-group {composed
 by}
  the LF boosts, which allows a separation of the intrinsic motion from the global one, very important for the 
 description of DIS, SIDIS and deeply virtual Compton scattering processes, since it is possible to unambiguously 
 identify the effects due to the inner dynamics.
 
Therefore our LF spectral functions can be useful in many problems that require both a proper relativistic treatment 
and at the same time a good description of the internal structure of the system. 

 As a first example of forthcoming applications, we can mention the study of the effect of relativity in the evaluation 
 of SIDIS cross section off $^3$He, taking into account both the relativity and the interaction in the final state 
 between the observed pion and the remnant. In Refs. \cite{Dotto1,tobe}, {by adopting}  a non relativistic spectral 
 function evaluated {from} the $^3$He wave function of Ref. \cite{pisa},  a distorted spin-dependent spectral function
  was obtained using a generalized eikonal approximation to deal with the final state interaction, 
  and it was shown that within this framework it is actually possible to get reliable information on the quark 
  TMDs in the neutron from SIDIS experiments off $^3$He. {By considering the new} LF spin-dependent spectral 
  function, we plan to  evaluate SIDIS cross sections off $^3$He  through {a LF} distorted spin-dependent spectral function obtained applying again the generalized eikonal approximation for the description of the final state interaction.
 Preliminary results can be found in Ref. \cite{Pace1}.
 
 A second example {for an application of the LF approch proposed in this paper}, is the study  of the role played by relativity in 
 the EMC effect on $^3$He, for which JLab data have been taken at 6 GeV \cite{Seely} in the standard inclusive DIS 
 sector. Encouraging  results including an exact treatment of the  deuteron channel and an approximated treatment for the continuum of the LF spectral function can be found in Ref. \cite{Rinaldi}.

 In view of the large efforts in the determination of the TMDs to study the three-dimensional structure of the nucleon,  the same concepts and definitions that are used in this paper to build up the LF spin-dependent spectral function for a three-nucleon system  could be tentatively applied to a  system of three valence quarks to define a nucleon spectral function in valence approximation and  then to describe the nucleon TMDs in terms of a valence wave function for the  nucleon.

It will be also interesting to study in detail the relation between the LF spin-dependent spectral function and  the correlator, $\Phi(k,P,S) $, of a nucleon of momentum $k$ 
  in a nucleus of momentum $P$ and spin polarization  $S$,
   defined in terms of the nucleon fields, in analogy to the quark correlator in a nucleon,
    defined in terms of the quark fields \cite{bdr}. In Refs. \cite{Dotto,Pace1} preliminary results were presented 
and it was shown that, in the valence approximation, a simple relation between the correlator
   and the LF spin-dependent spectral function naturally emerges and that 
   %within the LF approach    and in the valence approximations, 
   only three of the six time-reversal even TMDs at the leading twist \cite{bdr} are independent.  The relations among these TMDs could be experimentally checked to test our LF description of the spin-dependent spectral function.

\newpage

\appendix

\section{Two-body light-front wave function}
\label{appa}
{In this Appendix, some details {are given} on the two-body light-front wave function that are useful for the general discussion presented in Sect. \ref{lfd23}.}

\subsection{Completeness of two-body free states}
\label{appa1}
Let $\blf P$ be the total LF momentum for a two-particle system
\be
\blf P=\blf p_1 +\blf p_2  \quad .
\ee

The Jacobian from $\{{\blf p}_1,{\blf p}_2 \}$ to $\{{\blf P},\xi,{\bf k}_\perp \}$
is
\be
\left[{\partial ({\blf p}_1,{\blf p}_2)\over \partial ({\blf P},\xi,{\bf k}_\perp )}\right]=
P^+\ee
and the Jacobian from $\{{\blf p}_1,{\blf p}_2 \}$  to 
$\{{\blf P},k^+,{\bf k}_\perp \}$
is given by
{\be
\left[{\partial ({\blf p}_1,{\blf p}_2)\over \partial ({\blf P},k^+,{\bf k}_\perp) }
\right]=
{2  (1-\xi)\over M_0(1,2)}~ P^+ ={2 \xi (1-\xi)\over k^+}~ P^+\quad \quad , 
\ee}
with $M_0(1,2)$ defined by Eq. (\ref{M02b}), since
{\be
{\partial k^+ \over \partial \xi}= M_{0}(1,2) -\xi  ~{1
\over 2 M_0(1,2)}~{m^2
+|{\bf k}_{\perp}|^2\over \xi^2(1-\xi)^2}~(1-2 \xi)
=
{M_0(1,2)\over 2  (1-\xi)}={k^+\over 2 \xi (1-\xi)} \quad .
\label{der1}
\ee} 
Furthermore the Jacobian from $\{{\blf p}_1,{\blf p}_2 \}$  to 
$\{{\blf P},k_z,{\bf k}_\perp \}$
is given by 
\be
\left[{\partial ({\blf p}_1,{\blf p}_2 ) \over \partial ({\blf P},k_z,{\bf k}_\perp )}\right]=
{2 \xi (1-\xi)\over E({\bf k})}~ P^+  \quad \quad ,
\ee
since {(cf Eq. (\ref{kzz}))}
{\be
{\partial k_z \over \partial \xi}= M_{0}(1,2) -\left(\xi -{1\over 2}\right) ~{1
\over 2 M_{0}(1,2)}~{m^2
+|{\bf k}_{\perp}|^2\over \xi^2(1-\xi)^2}~(1-2 \xi)
={E({\bf k}) \over 2 \xi (1-\xi)} \quad \quad  .
\label{der2}
\ee}
From Eqs. (\ref{der1}) and (\ref{der2}) one has
\be
{\partial  k^+\over \partial  k_z}= {\partial k^+ \over \partial \xi} ~ {\partial  \xi  \over \partial k_z}
={k^+  \over E({\bf k})} \quad \quad  .
\label{der3}
\ee

Keeping separate the global motion from the intrinsic one, the completeness
 reads
\be
{\bf{I}}~=~  \int
{d{\blf  p}_1\over 2p^+_1(2\pi)^3}~{d{\blf  p}_2\over 2p^+_2(2\pi)^3} ~|{\blf p}_1\rangle |{\blf p}_2\rangle~
\langle{\blf p}_1|\langle{\blf p}_2|=
\nonu= 2~\int {d{\blf  P}\over 2P^+(2\pi)^3}|{\blf P}\rangle \langle{\blf P}|
\int {d\xi\over (2\pi)^3 ~4\xi (1-\xi)}
\int~d{\bf  k}_\perp ~
 {|{\blf k}\rangle~
\langle{\blf k}|}=
\nonu=
2~\int {d{\blf  P}\over 2P^+(2\pi)^3}|{\blf P}\rangle \langle{\blf P}|\int{ {d\blf  k}\over 2k^+(2\pi)^3 } ~
 |{\blf k}\rangle~
\langle{\blf k}|=
\nonu=
 2~\int {d{\blf  P}\over 2P^+(2\pi)^3}|{\blf P}\rangle \langle{\blf P}|
 \int {d{\bf  k}\over (2\pi)^3 2 E({\bf k})}~
  |{\blf k}\rangle~
\langle{\blf k}|  \quad .
\label{freecomp}
\ee
{Notice in the last step the hybrid notation in the intrinsic part. It will be used in what follows}.
%where $ |{\bf k}\rangle$ is the eigenstate corresponding to the intrinsic LF momentum ${\bf  k}$.

The normalization of the free state 
{$|{\blf P}\rangle  |{\blf k}\rangle~ = |{\blf p}_1\rangle  |{\blf p}_2\rangle$ is
\be
\langle {\blf p}'_2|{\blf p}_2\rangle~\langle {\blf p}'_1|{\blf p}_1\rangle
= 2p^+_1 (2\pi)^3~\delta^3({\blf p}'_1- {\blf p}_1 )~ 2p^+ _2(2\pi)^3~\delta^3({\blf p}'_2-
{\blf p}_2 )= \nonu=\left[{  \partial ({\blf
P},k^+,{\bf k}_\perp )\over \partial ({\blf p}_1,{\blf p}_2 )}\right]2p^+_1 (2\pi)^3~ 2p^+ _2(2\pi)^3~
\delta^3({\blf P}'- {\blf P} )~\delta^3({\blf k}'- {\blf k} )=\nonu=
2P^+ (2\pi)^3~\delta^3({\blf P}'- {\blf P} )~ k^+ (2\pi)^3~\delta^3({\blf k}'- {\blf k} )
=~\langle{\blf P}'|{\blf P}\rangle ~\langle{\blf k}' |{\blf k}\rangle  \quad .
\label{freenorm}\ee
It should be pointed out that 
$\langle{\blf k}' |{\blf k}\rangle =k^+ (2\pi)^3~\delta^3({\blf k}'- {\blf k} )$
, i.e. without a factor of two,
 since it refers to a two-body intrinsic state.}
%Moreover, the normalization of the free state $|{\blf P}\rangle  |{\bf k}\rangle$ is
%\be
%\langle{\blf P}'|{\blf P}\rangle ~\langle{\bf k}' |{\bf k}\rangle=
%2P^+ (2\pi)^3~\delta^3({\blf P}'- {\blf P} )~ E({\bf k}) (2\pi)^3~\delta^3({\bf k}'- {\bf k} )=\nonu=
%2P^+ (2\pi)^3~\delta^3({\blf P}'- {\blf P} )~ E({\bf k}) (2\pi)^3~\delta^2(\widehat{\bf k}'- \widehat{\bf k} )
% {\delta(k'-k) \over k^2}\label{freenorm1}\ee

{The overlap between the free two-body intrinsic states 
$|{\blf k};\sigma_2,\sigma_1\rangle_{LF}$ and the corresponding ones with canonical 
spin and Cartesian
coordinates is relevant for the following discussion. {Reminding  that 
$\delta(k^{\prime +} -k^+) = \delta(k'_z -k_z)/({\partial k^+ / 
\partial k_z})$ and } using Eq. (\ref{melosh}) one has 
\be
_c\langle \sigma'_1,\sigma'_2; {\bf k}' |{\blf k}; \sigma_2,\sigma_1\rangle_{LF}
=\sqrt{(2\pi)^3~k^+~{\partial k_z \over \partial k^+}}~
\delta({\bf k}' -{\bf k})~ D^{{1 \over 2}*} [{\cal R}_M ({\blf k} )]_{\sigma_1\sigma'_1}~
D^{{1 \over 2}*} [{\cal R}_M (-{\blf k} )]_{\sigma_2\sigma'_2}~~,
\label{melosh2}\ee
where the normalization and the completeness of the plane waves   with Cartesian variables,
$|{\bf k}\rangle$ are 
  \be
 \langle {\bf k'}| {\bf k}\rangle= \delta ({\bf k'} - {\bf k}) 
 \nonu
 \int {d{\bf  k} }~
  |{\bf k}\rangle~
\langle{\bf k}|~= ~{\bf{I}} \quad ,
  \ee
{and 
\be
-{\blf k}\equiv((M_0 - k^+),-{\bf k}_\perp) ~~.
\ee }
%{\Red Gia' detto sopra Notice the absence of a factor of 2  
%(cf Eq. (\ref{freenorm})), since the normalization is for the two-body case  \
%and not for a one-body state.}
\subsection{Light-front wave function for a system of two interacting particles}
\label{appa2}
{By using} the subgroup properties of the LF boosts, the LF  wave function for an interacting two-body system, in a given frame, can be expressed through the
intrinsic variables as follows {(see Eq. (\ref{melosh2}))}
\be
 _{LF}\langle \sigma_1,\sigma_2;  \tau_1,\tau_2;{\blf k},{\blf P}'
 |{\blf P};j,j_z;
 \epsilon_{int},\alpha; T T_z \rangle_{LF} 
 =2~P^+~
 (2\pi)^3~\delta ^3({\blf P}'  -{\blf P})~ {\sqrt {(2\pi)^3 k^+~\partial k_z/\partial k^+ }}
% \left[k^+~(2\pi)^3{\partial k_z \over \partial k^+}\right]^{1/2} 
  \nonu
 ~\times ~ \sum_{\sigma'_1,\sigma'_2}
~ D^{{1 \over 2}} [{\cal R}_M ({\blf k} )]_{\sigma_1\sigma'_1}~
D^{{1 \over 2}} [{\cal R}_M (-{\blf k} )]_{\sigma_2\sigma'_2}~
 %{\cal N}(\epsilon_{int})~
~\langle \sigma'_1,\sigma'_2;  \tau_1,  
\tau_2; {\bf k}|j,j_z;
\epsilon_{int},\alpha; T T_z 
 \rangle~~~~,
\label{lfwf2}\ee 
%where LF and Cartesian momenta are properly considered and 
%$$\left[k^+~(2\pi)^3{\partial k_z \over \partial
% k^+}\right]^{1/2}= \sqrt {(2\pi)^3 E^{in}({\bf k}) }$$
where  a canonical completeness has been inserted for obtaining the final 
step.
% {\Red (cf Eq. (\ref{melosh2})}.

 Notice that    the {intrinsic two-body wave function} 
$\langle \sigma'_1,\sigma'_2;  \tau_1,  \tau_2; {\bf k}|j,j_z; \epsilon_{int},\alpha; T T_z  \rangle $ 
 %$$\delta ({|{\bf k}|^2 \over M} -\epsilon_{int})$$
% Moreover it 
 contains  {\em canonical spins}, and therefore it can be {composed} by using the Clebsh-Gordan
 coefficients. Moreover, $j$ is the total angular momentum of the pair, 
 %(as defined within the Bakamijan-Thomas approach), 
 $T$  the isospin, $\alpha$  the set of the
  parity and  quantum numbers that label the coupled waves, and 
 $\epsilon_{int}$ is the eigenvalue of the mass operator 
 (see Eqs. (\ref{eigen2},\ref{eigen3},\ref{eigen4})). 

The normalization of the {\em intrinsic part} of a LF {\em bound  state}
follows from  the normalization
fulfilled by $\langle \sigma_1,\sigma_2; \tau_1,  
\tau_2; {\bf k}|j,j_z;\epsilon_{int},\alpha ;T T_z\rangle $. 
Indeed, if we adopt the
following normalization, suitable for bound states,
 \be \sum_{\tau_1,\tau_2}
 \sum_{\sigma_1,\sigma_2}~
  \int {d{\bf k} } ~|\langle \sigma_1,\sigma_2; 
  \tau_1, \tau_2; {\bf k}|j,j_z;\epsilon_{int},\alpha ; T T_z
 \rangle|^2~=~1~~~~~~,
 %{\bf{I}} 
 \ee
 from Eq. (\ref{lfwf2}) one has for the intrinsic part of the two-body LF wave function
 \be \sum_{\tau_1,\tau_2}
 \sum_{\sigma_1,\sigma_2}
\int {d k^+ d{\bf k}_\perp\over k^+ (2\pi)^3} 
~|_{LF}\langle \sigma_1,\sigma_2;\tau_1,\tau_2;
{\blf k}|j,j_z;
 \epsilon_{int},\alpha; T T_z \rangle|^2=
 \nonu = 
\sum_{\tau_1,\tau_2}
 \sum_{\sigma_1,\sigma_2}
\int {d{\bf k}\over E({\bf k}) (2\pi)^3} 
~|_{LF}\langle \sigma_1,\sigma_2;\tau_1,\tau_2;
{\blf k}|j,j_z;
 \epsilon_{int},\alpha; T T_z \rangle |^2=
 \nonu =   
 \sum_{\tau_1,\tau_2}
 \sum_{\sigma_1,\sigma_2}~
 ~\int {d{\bf k} \over E({\bf k})}~
  {E({\bf k})}
\left | \sum_{\sigma'_1,\sigma'_2} D^{{1 \over 2}} [{\cal R}_M({\blf k} )]_{\sigma_1\sigma'_1}~
D^{{1 \over 2}} [{\cal R}_M (-{\blf k} )]_{\sigma_2\sigma'_2}    ~
%\right . ~\nonu \times ~  \left .
 \langle \sigma'_1,\sigma'_2; \tau_1,\tau_2;{\bf k}|j,j_z;
 \epsilon_{int},\alpha; T T_z \rangle \right  | ^2=\nonu=
 ~\sum_{\tau_1,\tau_2}
 \sum_{\sigma_1,\sigma_2}
 ~\int d{\bf k}  ~ 
|\langle \sigma_1,\sigma_2; \tau_1,\tau_2; {\bf k}|j,j_z;
\epsilon_{int},\alpha ; T T_z
 \rangle|^2=~1 ~~.
 %{\bf{I}}
 \label{norma2}
 \ee
{In the last step of (\ref{norma2}) the unitarity of the  $D^{1/2}$ matrices has been used.}

The normalization 
%factor 
for the LF {\em scattering states} follows from : {(i) the orthogonality} condition
adopted for the canonical scattering wave function  
$\langle \sigma_1,\sigma_2; \tau_1,\tau_2; {\bf k}|j,j_z; \epsilon_{int},\alpha ; T T_z \rangle$,
 given by (see also Eq. (\ref{nrcompl}) below for the completeness of the canonical states) 
%one can imposes 9notice that one can recover the completenes $\int
% d{\bf k}/E^{in}$ through a suitable redefintion of the overlap), viz
 \be
 \sum_{\sigma''_1,\sigma''_2}\sum_{ \tau''_1,\tau''_2} \int d {\bf k}~
 \langle T'_z T'; \alpha'\epsilon'_{int};j'_zj'|
 {\bf k};\tau''_2,\tau''_1 ;\sigma''_2,\sigma''_1
 \rangle ~\langle \sigma''_1,\sigma''_2;  \tau''_1,\tau''_2;{\bf k}
 |j,j_z;
 \epsilon_{int},\alpha; T T_z \rangle
 =\nonu =\delta_{T',T} ~
\delta_{T'_z,T_z} \delta_{\alpha',\alpha} \delta_{j',j}
\delta_{j'_z,j_z}~~{\delta (t'-t)\over t^2}  \quad ,
 \ee 
 where $t=\sqrt{m \epsilon_{int}}$, and {(ii) the orthogonality} adopted for the LF scattering states, that reads
{(see also the completeness of the free states for a two-body system $|{\blf P}\rangle|{\blf k}\rangle$  in   Eq. (\ref{freecomp}))},
\be
_{LF}\langle T'_z T';\alpha' \epsilon'_{int} j'_z j';{\blf P}'|{\blf P};j,j_z;
 \epsilon_{int},\alpha; T T_z \rangle_{LF}=
 \nonu
 =  \sum_{\sigma''_1,\sigma''_2}\sum_{ \tau''_1,\tau''_2} \int {d {\blf
 P}''\over 
 2 P^{\prime \prime +} (2 \pi)^3} \int {d {\bf
 k}\over E({\bf k})(2 \pi)^3}
\nonu 
 _{LF}\langle T'_z T'; \alpha'\epsilon'_{int};j'_zj';{\blf P}'|
 {\blf P}'',{\blf k};\tau''_2,\tau''_1 ;\sigma''_2,\sigma''_1
 \rangle_{LF} ~~_{LF}\langle \sigma''_1,\sigma''_2;  \tau''_1,\tau''_2;{\blf k},{\blf P}''
 |{\blf P};j,j_z;
 \epsilon_{int},\alpha; T T_z \rangle_{LF}=\nonu
 = 2~P^+~(2\pi)^3~\delta ^3({\blf P}'  -{\blf P}) 
% (2\pi)^3~ (m^2+m\epsilon'_{int})^{1/4}~(m^2+m\epsilon_{int})^{1/4}
 \sum_{\sigma''_1,\sigma''_2}\sum_{ \tau''_1,\tau''_2} \int d {\bf k}
\nonu 
 \langle T'_z T'; \alpha'\epsilon'_{int};j'_zj'|
 {\bf k};\tau''_2,\tau''_1 ;\sigma''_2,\sigma''_1
 \rangle ~\langle \sigma''_1,\sigma''_2;  \tau''_1,\tau''_2;{\bf k}
 |j,j_z;
 \epsilon_{int},\alpha; T T_z \rangle=
 \nonu=
 2~P^+~(2\pi)^3~\delta ^3({\blf P}'  -{\blf P})~\delta_{T',T} ~
\delta_{T'_z,T_z} \delta_{\alpha',\alpha} \delta_{j',j}
\delta_{j'_z,j_z}~
%(2\pi)^3~\sqrt{m^2+m\epsilon_{int}}
~{\delta (t'-t)\over t^2}  \quad .
\label{freeStateOrt}
\ee

Then for the two-body interacting case the LF completeness reads
\be
\int {d{\blf P} \over 2 P^+ (2 \pi)^3} \sum_{j,j_z\alpha} ~ \sum_{T T_z}~
 \sumint {\lambda(t) ~ dt
% (2\pi)^3 \sqrt{m^2+m\epsilon_{int}}
 }
~_{LF} \langle \sigma_1,\sigma_2;  \tau_1,\tau_2;{\blf k},{\blf P}'|
{\blf P};j,j_z; \epsilon_{int},\alpha; T T_z \rangle_{LF} \nonu
\times ~_{LF}\langle T_z T;\alpha,\epsilon_{int};j_z,j;{\blf P}|
 {\blf P}'',{\blf k}';\tau'_2,\tau'_1 ;\sigma'_2,\sigma'_1 \rangle _{LF}
  \nonu
 =~  2~P'^+~(2\pi)^3~\delta ^3({\blf P}'  -{\blf P}'')~ \sum_{j,j_z\alpha} \sum_{T T_z}~
 \sumint {\lambda(t) ~ dt 
}
% \left[k^+~(2\pi)^3{\partial k_z \over \partial k^+}\right]^{1/2} 
  \nonu \times
~
 \sqrt {(2\pi)^3 E({\bf k}) }
 \sum_{\bar{\sigma}_1,\bar{\sigma}_2}
~ D^{{1 \over 2}} [{\cal R}_M({\blf k} )]_{\sigma_1\bar{\sigma}_1}~
D^{{1 \over 2}} [{\cal R}_M (-{\blf k} )]_{\sigma_2\bar{\sigma}_2}~
~\langle \bar{\sigma}_1,\bar{\sigma}_2;  \tau_1,  
\tau_2; {\bf k}|j,j_z;
\epsilon_{int},\alpha; T T_z  \rangle 
% \left[k^+~(2\pi)^3{\partial k_z \over \partial k^+}\right]^{1/2} 
 \nonu
 ~\times ~
 \sqrt {(2\pi)^3 E({\bf k'}) }
 \sum_{\bar{\sigma}'_1,\bar{\sigma}'_2}
~ D^{{1 \over 2}\dagger} [{\cal R}_M ({\blf k'} )]_{\bar{\sigma}'_1\sigma'_1}~
D^{{1 \over 2}\dagger} [{\cal R}_M (-{\blf k'} )]_{\bar{\sigma}'_2\sigma'_2}~
~\langle j,j_z;
\epsilon_{int},\alpha; T T_z 
|\bar{ \sigma}'_1,\bar{\sigma}'_2;  \tau_1',  \tau_2'; {\bf k'}\rangle ~
   \nonu
=~  2~P'^+~(2\pi)^3~\delta ^3({\blf P}'  -{\blf P}'')~\delta_{\tau_1',\tau_1} 
\delta_{\tau_2',\tau_2} \delta_{\sigma_1',\sigma_1} \delta_{\sigma_2',\sigma_2}
\delta ^3({{\blf k}'  -{\blf k}})~(2\pi)^3 ~ k^+  \quad ,
\label{tbcompl}
\ee 
where the symbol $\int \! \! \! \!\! \!\sum$ means a sum over the
bound states of the  pair (namely the deuteron in the present case) 
and the integration over the 
continuum. The quantity $\lambda(t)$ is the $t$-density of the two-body states 
($\lambda(t)= 1$ for the bound states and 
$\lambda(t)= t^2 $ for the continuum). To obtain Eq.  (\ref{tbcompl}), 
one has to use : {(i) the expression  (\ref{lfwf2}) for the LF wave function, 
(ii) 
the unitarity of the $D^{1/2}$ matrices, (iii) the completeness for the
eigensolutions of Eq. (\ref{eigen2}), i.e.,
\be
\sum_{j,j_z\alpha} \sum_{T T_z}\sumint {\lambda(t) ~  dt}
\langle{\bf k}' |j,j_z; \epsilon_{int},\alpha; T T_z \rangle ~
\langle T_z T;\alpha,\epsilon_{int};j_z,j|{\bf k}\rangle=
\delta^3({\bf k}'-{\bf k}) \quad \quad  ,
\label{nrcompl}
\ee 
and (iv)} Eq. (\ref{der3}). 
%Summarizing, the above discussion strongly suggests to approximate the overlap $\langle \sigma_1,\sigma_2; 
 % \tau_1, \tau_2; {\bf k}|j,j_z;\epsilon_{int},\alpha ; T T_z
 %\rangle$ with  the {\em non relativistic  counterpart}. 

%f we go back to the light-cone variables $\{\xi,{\bf k}_\perp \}$ in order to
%have light-cone distributions for constituents inside a two-body bound state, 
%one can write
%\be
%n(\xi,|{\bf k}_\perp|)= ~{1\over 2\xi (1-\xi)(2\pi)^3}~
%\sum_{\sigma_1\sigma_2}\sum_{\tau_1\tau_2}~|_{LF}\langle \sigma_1,\sigma_2; \tau_1,\tau_2; {\bf k}|j,j_z;
%\epsilon_{int},\alpha ; T T_z \rangle|^2 \ee
 %With the normalization that follows from= Eq. (\ref{norma2})
%\be
%\int d\xi \int d{\bf k}_\perp~n(\xi,|{\bf k}_\perp|)=1
%\ee
%The momentum sum rule can be easily obtained, by recalling that for particle
%''1'' one has $\xi=k^+_1/M_0$, then it follows
%\be
%\int d\xi \int d{\bf k}_\perp~\xi~n(\xi,|{\bf k}_\perp|)=
%\int d{\bf k}_1  ~ {k^+_1 \over M_0}
%|\langle \sigma_1,\sigma_2; \tau_1,\tau_2; {\bf k}_1|j,j_z;
%\epsilon_{int},\alpha ; T T_z
% \rangle|^2= \nonu= \int d{\bf k}_1 \int d{\bf k}_2 ~
% \delta^3({\bf k}_1+{\bf k}_2)~ {k^+_1 \over M_0}
%|\langle \sigma_1,\sigma_2; \tau_1,\tau_2; {\bf k}_1|j,j_z;
%\epsilon_{int},\alpha ; T T_z
% \rangle|^2= {1 \over 2}
%\ee
%The result follows from the symmetry of the two-body three-dimensional wave function and from
%$k^+_1+k^+_2=M_0$ (kinematical nature of the plus component). 

\section{Three-body  states }
\label{appb}
{In this Appendix, the three-body free and interacting states are analyzed in analogy to the two-body case.}
\subsection{Completeness of three-body free states with symmetric intrinsic variables}
\label{appb1}
Let $\blf P$ be the total LF momentum for a three-particle system
\be
\blf P=\blf p_1 +\blf p_2+\blf p_3
\ee
of free mass $M_0(1,2,3)$
\be
M^2_0(1,2,3)={m^2+|{\bf k}_{1\perp}|^2\over \xi_1}+{m^2+|{\bf k}_{2\perp}|^2\over \xi_2}
+{m^2+|{\bf k}_{3\perp}|^2\over \xi_3}=(E_1+E_2+E_3)^2  \quad \quad ,
\ee
where $E_i=\sqrt{m^2+|{\bf k}_{i}|^2}$ and $\sum_i{\bf k}_{i}=0$.
 
 The completeness  for the different
set of variables, 
$\{{\blf p}_i\}\to
\{\xi_i,{\bf k}_{i\perp}\}\to {\bf k}_i
%\to{\blf k}_i
$, is given by
\be
{\bf I} ~ = ~\int ~{d {\blf p}_1 
\over 2 p^+_1 (2 \pi)^3}{d {\blf p}_2 \over 2 p^+_2(2 \pi)^3}
{d {\blf p}_3 \over 2 p^+_3(2 \pi)^3}~|{\blf p}_3\rangle
|{\blf p}_2\rangle |{\blf p}_1\rangle \langle{\blf p}_1|\langle{\blf p}_2|
\langle{\blf p}_3|
=\nonu=
 \int {d {\blf P} \over 2 P^+ (2 \pi)^3}~|{\blf P}\rangle \langle{\blf P}|~
 \int{d\xi_1\over 2 \xi_1(2 \pi)^3}~d {\bf k}_{1 \perp} 
 |\xi_1 {\bf k}_{1 \perp} \rangle \langle{\bf k}_{1 \perp} \xi_1|
\int {d\xi_2\over 2 \xi_2(2 \pi)^3}~d {\bf k}_{2 \perp}{1\over  \xi_3}~
|\xi_2 {\bf k}_{2 \perp} \rangle \langle{\bf k}_{2 \perp} \xi_2| =\nonu=
\int {d {\blf P} \over 2 P^+(2 \pi)^3}~|{\blf P}\rangle \langle{\blf P}|
\int {d {\bf k}_{1}\over 2 E_1(2 \pi)^3} 
~\int {d {\bf k}_{2 }\over 2 E_2(2 \pi)^3}~{M_0(1,2,3)\over  E_3}
|{\blf k}_1\rangle|{\blf k}_2 \rangle
 \langle{\blf k}_2|\langle{\blf k}_1| \quad ,
%=\nonu=
%\int {d {\blf P} \over P^+(2 \pi)^3}~|{\blf P}\rangle \langle{\blf P}|~\int
%{d {\blf k}_{1 }\over 2 k^+_1(2 \pi)^3} \int {d {\blf k}_{2 }\over
%2 k^+_2(2 \pi)^3}~{M_0(1,2,3)\over  k^+_3}|{\blf k}_1\rangle|{\blf k}_2 \rangle
 %\langle{\blf k}_2|\langle{\blf k}_1|
 \label{comp3}
 \ee
 where $ |{\blf p}_3\rangle|{\blf p}_2\rangle |{\blf p}_1\rangle = 
 |{\blf P}\rangle |{\blf k}_1\rangle|{\blf k}_2 \rangle =
  |{\blf P}\rangle |\xi_1,{\bf k}_{1\perp}\rangle|\xi_2,{\bf k}_{2\perp} \rangle $ 
 and the Jacobians
 \be
\left [ {\partial ({\blf p}_1,{\blf p}_2,{\blf p}_3)\over 
\partial ({\blf P},\xi_1,{\bf k}_{1\perp}, \xi_2,{\bf k}_{2\perp})}\right ] =
(P^+)^2
 \label{Jac}
 \ee
  \be
\left [ {\partial ({\blf p}_1,{\blf p}_2,{\blf p}_3)\over 
\partial ({\blf P},{\bf k}_{1},{\bf k}_{2})}\right ] =
{p_1^+ ~ p_2^+ ~ p_3^+ M_0(1,2,3) \over P^+ E_1 ~ E_2 ~ E_3} 
 \label{Jac1}
 \ee
 have been used.
 
%In the intrinsic frame of the three-particle system,  
%instead of using
 %$\{\xi_i,{\bf k}_{i\perp}\}$ one can use
%$\{k_{iz},{\bf k}_{i\perp}\}$.
%Then, one can introduce a set of  Jacobi momenta (Cartesian momenta) 
%and conjugate coordinates as follows
%\be {\bf p }= { 2 \over 3} \left (
%{\bf k}_1 - {{\bf k}_2 + {\bf k}_3 \over 2} \right )
%= {\bf k}_1 \quad \quad \rhobf= \left (
%{\bf r}_1 - {{\bf r}_2 + {\bf r}_3 \over 2} \right )
%\nonu {\bf k} = { 1\over 2} \left({\bf k}_2 -{\bf k}_3\right ) = 
%{\bf k}_2 + { {\bf k}_1\over 2} \quad \quad ~{\bf r}= \left({\bf r}_2 -{\bf r}_3 \right ) 
%\ee 
%Notice that the Jacobian of the transformation is 1 for both momenta and coordinates.

\subsection{Completeness of three-body free states with non-symmetric intrinsic variables} 
\label{appb2}
Instead of the symmetric intrinsic variables in the
3-body frame, one can introduce non-symmetric intrinsic
variables, corresponding to the intrinsic frame of the (2,3) pair, i.e.
$\{{\blf p}_2,{\blf p}_3\} \to \{{\blf P}_{23},\eta, {\bf k}_{23\perp}\}$ 
(see Eqs. (\ref{nonsymv},\ref{nonsymv1})). 

The completeness 
\be
\int
{d{\blf  p}_1\over 2p^+_1(2\pi)^3}~{d{\blf  p}_2\over 2p^+_2(2\pi)^3} ~{d{\blf 
p}_3\over 2p^+_3(2\pi)^3} |{\blf p}_1\rangle|{\blf p}_2\rangle |{\blf p}_3\rangle~
\langle{\blf p}_3|\langle{\blf p}_2|~
\langle{\blf p}_1| ~= ~{\bf \rm I}
\label{simcompl}\ee
can be arranged in different ways, depending upon the the choice of variables
one needs.
In particular,  
\begin{enumerate}
\item for the variables ${\blf  p}_1$, ${\blf  P}_{23}$ and ${\blf k}_{23}$
 one can
exploit Eq. (\ref{freecomp}), obtaining
\be 
 {\bf \rm I} =\int
{d{\blf  p}_1\over 2p^+_1(2\pi)^3}~|{\blf p}_1\rangle
\langle{\blf p}_1| \int{d{\blf  P}_{23}\over 2P^+_{23}(2\pi)^3} ~
|{\blf P}_{23}\rangle~
\langle{\blf P}_{23}|\int {d{\blf k}_{23}\over k_{23}^+(2\pi)^3}~ |{\blf k}_{23}\rangle 
\langle{\blf k}_{23}|
\label{compl1}\ee
\item for the variables ${\blf  P}$, $\{\xi_1,{\bf k}_{1\perp}\} $
 and $\{ \eta,{\bf k_{23}}_{\perp}\}$ one has from Eq. (\ref{comp3})
\be
 {\bf \rm I} =
  \int {d{\blf  P}\over 2P^+(2\pi)^3}~|{\blf P}\rangle\langle{\blf P}|\int {d\xi_1~d{\bf  k}_{1\perp}
  \over 2 \xi_1 (1-\xi_1)
(2\pi)^3 }~| \xi_1{\bf k}_{1\perp}\rangle\langle{\bf k}_{1 \perp} \xi_1|
~\nonu \times ~\int {d\eta~d{\bf  k}_{23\perp}\over 2 \eta (1-\eta)(2\pi)^3 }
~|\eta {\bf k}_{23\perp}\rangle~
\langle{\bf k}_{23\perp} \eta| \quad \quad ,
\label{compl2}\ee
after recalling Eq. (\ref{nonsymv}) that yields
\be
{d \xi_2\over \xi_2\xi_3}={d \eta\over \eta (1-\eta) (1-\xi_1)} \quad 
 {\rm and} \quad d {\bf k}_{2\perp}=d {\bf k}_{23\perp}  
 \ee
\item for the variables ${\blf  P}$, ${\blf k}_{1} $
 and ${\blf  k}_{23}$ one has
\be
{\bf \rm I} =\int {d{\blf  P}\over 2P^+(2\pi)^3}~|{\blf P}\rangle\langle{\blf P}|
\int {d{\blf  k}_{23}\over  k_{23}^+
(2\pi)^3 }| {\blf k}_{23}\rangle~\langle{\blf k}_{23}|
~\int {M_0(1,2,3) ~d{\blf  k}_1\over 2k^+_1 E_{23}(2\pi)^3 }
~
 | {\blf k}_{1}\rangle 
\langle{\blf k}_1  | \quad ,
\label{compl3}\ee
where the following relations have been used (remind that $k^+_1=\xi_1M_0(1,2,3)$ and $k_{23}^+=\eta M_{23}$)
\be
{\partial k^+_1 \over \partial \xi_1}= M_0(1,2,3) + \xi_1 {\partial M_0(1,2,3) \over \partial \xi_1}=
M_0(1,2,3) + {\xi_1 \over 2 M_0(1,2,3)}{\partial M^2_0(1,2,3) \over \partial \xi_1}=\nonu=
{1 \over 2M_0(1,2,3)}~\left[M^2_0(1,2,3)
+{M^2_{23}+|{\bf k}_{1\perp}|^2\over (1-\xi_1)^2}\right]=
{1\over 2 (1-\xi_1)}~\left[M_0(1,2,3) (1-\xi_1)
+{M^2_{23}+|{\bf k}_{1\perp}|^2\over {K}^{+}_{23}}\right]=\nonu=
{1 \over 2 (1 -\xi_1)} ~\left[{K}^+_{23}+ {K}^-_{23on }\right]={E_{23}\over(1-\xi_1)}
\label{deracompl3}\\ &&
\nonu
{\partial k_{23}^+ \over \partial \eta}=  M_{23} -\eta  ~{1
\over 2 M_{23}}~{m^2
+|{\bf k}_{1\perp}|^2\over \eta^2(1-\eta)^2}~(1-2 \eta)
=\nonu={M_{23}\over 2 (1-\eta)}~\left [ 2 (1-\eta)-1+2\eta)\right]
={M_{23} \over 2 (1-\eta)}= {k_{23}^+\over 2 \eta (1-\eta)} 
\label{derbcompl3}\ee
with ${K}_{23}$ the total momentum of the free (2,3) pair in the intrinsic frame of the three
particles. i.e. ${K}_{23}^+ =  M_0(1,2,3)(1-\xi_1)$, 
${K}_{23\perp} = {\bf k}_{2\perp} + {\bf k}_{3\perp} = -{\bf k}_{1\perp}$, 
${K}_{23on}^- = (M^2_{23}+|{\bf k}_{1\perp}|^2 )/ {K}^+_{23}$,
 and $E_{23} = \sqrt{M_{23}^2 + |{\bf k}_{1}|^2}$.

\item for the variables ${\blf  P}$, ${\bf k}_{1} $
 and ${\bf  k_{23}}$ one has
\be
{\bf \rm I} =\int {d{\blf  P}\over 2P^+(2\pi)^3}~|{\blf P}\rangle\langle{\blf P}|
\int {2~d{\bf   k}_{23}\over M_{23} (2\pi)^3 }|{\blf k}_{23}\rangle~\langle{\blf  k}_{23}|
~\int {M_0(1,2,3)~d{\bf  k}_1\over 2 E_1 E_{23}(2\pi)^3 }
~
 |{\blf k}_1\rangle 
\langle{\blf k}_1| \quad .
\label{compl4}
\ee

For obtaining the above results, the following properties have been used 
\be
{\partial k_{1z} \over \partial k^+_1}={1 \over 2} \left[1 + {m^2+|{\bf k}_{1\perp}|^2\over
k^{+2}_1}\right]={ E({\bf {k}_1})\over k^+_1}={ E({\bf {k}_1}) \over M_0(1,2,3)\xi_1}
\label{deracompl4}\\  &&
\nonu
\nonu
{\partial k_{23z} \over \partial \eta}= M_{23} -\left(\eta -{1\over 2}\right) ~{1
\over 2 M_{23}}~{m^2
+|{\bf k}_{1\perp}|^2\over \eta^2(1-\eta)^2}~(1-2 \eta)
=\nonu={M_{23}\over 4 \eta (1-\eta)}~\left [ 4\eta (1-\eta)+(2\eta-1))^2\right]
={M_{23} \over 4 \eta (1-\eta)}
\label{derbcompl5}
\ee

\item for the variables ${\blf  P}$, ${\blf k}_{1} $
 and ${\bf k_{23}}$ one has
 \be
{\bf \rm I} =\int {d{\blf  P}\over 2P^+(2\pi)^3}~|{\blf P}\rangle\langle{\blf P}|
\int {2~d{\bf  k_{23}}\over M_{23} (2\pi)^3 }|{\blf k}_{23}\rangle~\langle{\blf  k}_{23}|
~\int {M_0(1,2,3)~d{\blf  k}_1\over 2k^+_1 E_{23}(2\pi)^3 }
~
 |{\blf k}_1\rangle 
\langle{\blf k}_1| \quad .
\label{compl5}\ee
\end{enumerate}

\subsection{Useful derivatives involving non-symmetric intrinsic variables}
\label{appb3}
Let us  evaluate  the derivatives ${\partial \kappa^+_1 /  \partial \xi_1}$ and 
${\partial \kappa_{1z}  / \partial \kappa^+_1}$ :
\be
{\partial \kappa^+_1 \over \partial \xi_1}= {\cal{M}}_0(1,23) + \xi_1 
{\partial {\cal{M}}_0(1,23) \over \partial \xi_1}=
{\cal{M}}_0(1,23)  + {\xi_1 \over 2 {\cal{M}}_0(1,23)}{\partial {\cal{M}}_0(1,23)^2 \over \partial \xi_1}=\nonu=
{1 \over 2{\cal{M}}_0(1,23)}~\left[{\cal{M}}_0(1,23)^2
+{M^2_S+|{\bf k}_{1\perp}|^2\over (1-\xi_1)^2}\right]=
{1\over 2 (1-\xi_1)}~\left[{\cal{M}}_0(1,23) (1-\xi_1)
+{M^2_S+|{\bf k}_{1\perp}|^2\over P^{+}_S}\right]=\nonu=
{1 \over 2 (1 -\xi_1)} ~\left[P^+_S+ P^-_{ S on}\right]={E_S\over(1-\xi_1)}
\label{devkappa}
 \ee
%with $E_S = \sqrt{M_S^2 + |{\bf \kappa}|^2}$.

\be
{\partial \kappa_{1z} \over \partial \kappa^+_1}={1 \over 2} \left[1 + {m^2+|{\bf k}_{1\perp}|^2\over
\kappa^{+2}_1}\right]={  E({\bf  \kappa}_1) \over \kappa^+_1}=
{ E({\bf  \kappa}_1) \over {{\cal {M}}_0(1,23)\xi_1}} ~ \quad \quad .
\label{devkappa1}
\ee

\subsection{Normalization of the light-front wave function}
\label{appb4}
{Let us check that the factors
in  the expression of the intrinsic part of the LF wave function given by the second and the third 
lines of Eq. (\ref{lfwf4}))  allow one
 to obtain the normalization of the bound state
  $| j,j_z; \epsilon^3_{int},\Pi; {1 \over 2}, T_z \rangle $.}
Indeed using Eqs. (\ref{compl3}) and (\ref{compl4})) one has
\be
{\langle T_z {1\over 2};\Pi ,\epsilon^3_{int}; j_z,
j|j,j_z;\epsilon^3_{int},\Pi ; {1\over 2} T_z \rangle}~ = \nonu
= \sum_{\tau_1,\tau_2,\tau_3}\sum_{\sigma_1,\sigma_2,\sigma_3}
\int {d{\blf  k}_{1}\over 2 k^+_1
(2\pi)^3 }
~\int {M_0(1,2,3)~d{\blf  k}_{23}\over  k^{+}_{23}~E_{23}(2\pi)^3 }~
| _{LF}\langle \sigma_1,\sigma_2,\sigma_3;  \tau_1,\tau_2,\tau_3; 
{\blf k}_1,{\blf k}_{23}|  j,j_z;
 \epsilon^3_{int},\Pi; {1 \over 2}, T_z \rangle |^2= \nonu
 = \sum_{\tau_1,\tau_2,\tau_3}
\sum_{\sigma_1\sigma_2\sigma_3}
 \int {d{\bf k}_1 \over  E_1 (2 \pi)^3} 
 \int {d{\bf  k}_{23}\over E_{23} } {M_0(1,2,3) \over (2 \pi)^3 M_{23}} ~
 |_{LF}\langle \sigma_1,\sigma_2,\sigma_3; \tau_1, \tau_2,\tau_3; {\blf k}_1,{\blf k}_{23}|j,j_z;
\epsilon^3_{int},\Pi ; {1\over 2} T_z
 \rangle|^2=
 \nonu=
\sum_{\tau_1,\tau_2,\tau_3} \sum_{\sigma_1,\sigma_2,\sigma_3}
\int {d{\bf  k}_{1}\over  E_1
(2\pi)^3 }
~\int {M_0(1,2,3)~d{\bf  k}_{23}\over M_{23} E_{23}(2\pi)^3 }~{2 E_1 M_{23}E_{23}(2\pi)^6
\over
  2 M_0(1,2,3)}~\nonu \times ~
| \sum_{\sigma'_1}\sum_{\sigma'_2}\sum_{\sigma'_3}~
D^{{1 \over 2}} [{\cal R}_M ({\blf k}_1 )]_{\sigma_1\sigma'_1}~
D^{{1 \over 2}} [{\cal R}_M ({\blf k}_{2} )]_{\sigma_2\sigma'_2}~ 
D^{{1 \over 2}}
 [{\cal R}_M( {\blf k}_{3} )]_{\sigma_3\sigma'_3}~\nonu \times ~
   \langle \sigma'_1,\sigma'_2,\sigma'_3;   \tau_1,\tau_2,\tau_3;
{\bf k}_1,{\bf  k}_{23}|  j,j_z;
 \epsilon^3_{int},\Pi; {1 \over 2}, T_z \rangle |^2=\nonu=
\sum_{\tau_1,\tau_2,\tau_3}\sum_{\sigma_1,\sigma_2,\sigma_3}
\int d{\bf  k}_{1}
~\int {d{\bf  k}_{23}}~
| \langle \sigma_1,\sigma_2,\sigma_3;   \tau_1,\tau_2,\tau_3;
{\bf k}_1,{\bf  k}_{23}|  j,j_z;
 \epsilon^3_{int},\Pi; {1 \over 2}, T_z \rangle |^2 = 1 \quad ,
\label{normhe3}
\ee
given the unitarity of the Melosh rotations and the normalization of the canonical wave function 
(\ref{norm3}).

\section{Properties of  the  basis states of the cluster $\left\{1,(2 3)\right\}$ }
\label{appc}
{In this Appendix, the general formalism, suitable for describing the cluster $\{1,(2 3)\}$, is presented. It should be reminded that the
 final goal is to
construct
 states where the  interaction is
acting only between the particles $2$ and $3$, namely the three-body states  we are
 {interested} in are the tensor product of  free one-body states and 
 interacting two-body states.}

\subsection{Completeness relation  for the non-symmetric basis states and orthogonality properties of three-body free states}
\label{appc1}
{The correctness of the normalization factors in Eq. (\ref{lfwf3}) can be 
 checked as follows.}

Indeed, let us consider the product of two three-body free states :
\be
A  = ~ _{LF}\langle \sigma'_1,\sigma'_2,\sigma'_3;\tau'_1,\tau'_2,\tau'_3;{\blf P}',{\blf k}'_1,{\blf k}'_{23}   |{\blf k}''_{23},{\blf k}''_1,{\blf P}'';\tau''_1,\tau''_2,\tau''_3; \sigma''_1,\sigma''_2,\sigma''_3\rangle_{LF} 
\quad \quad  .
\label{orto}
\ee
Then, let us insert in Eq. (\ref{orto}) the completeness relation (\ref{complns}) for the non-symmetric basis states   (\ref{1e23})
\be
A =  \int {d{\blf  P}\over 2P^+(2\pi)^3}\sum_{\sigma_1 \tau_1} 
\int {d\tilde{\bm \kappa}_{1}\over 2 \kappa^+_1 (2\pi)^3}
~\sum_{T_{23} \tau_{23}}\sumint {\lambda(t)~d{  t} 
%\over {\cal N}(\epsilon_{23})^2
}  
%{2 M_0 \over M_{23}E_{23}(2 \pi)^3}
~\nonu \times ~
\sum_{j_{23}j_{z23} \alpha}
~_{LF}\langle \sigma'_1,\sigma'_2,\sigma'_3;\tau'_1,\tau'_2,\tau'_3;{\blf P}',{\blf k}'_1,{\blf k}'_{23}   
|{\blf P}; \tilde{\bm \kappa}_1 \sigma_1 \tau_1; j_{23},
j_{z23}; \epsilon_{23},\alpha; T_{23}, \tau_{23} \rangle_{LF}~  
 \nonu 
\times ~ _{LF} \langle T_{23}, \tau_{23}; \alpha,\epsilon_{23};j_{z23},j_{23}; \tau_1 \sigma_1 \tilde{\bm \kappa}_1;{\blf P}|{\blf k}''_{23},{\blf k}''_1,{\blf P}'';\tau''_1,\tau''_2,\tau''_3;
 \sigma''_1,\sigma''_2,\sigma''_3\rangle_{LF} \quad .
\label{orto1}
\ee

With the help of the overlap {in  Eq. (\ref{lfwf3})}, the above equation  reads
\be
A ~
=~2~P'^+~
 (2\pi)^3~\delta ^3({\blf P}'' -{\blf P'}) ~ \delta_{\tau'_1\tau''_1}~\sum_{\sigma_1} 
\int {d\tilde{\bm \kappa}_{1}\over 2 \kappa^+_1 (2\pi)^3}
~\sum_{T_{23} \tau_{23}}\sumint {\lambda(t)~d{  t} 
} 
~\nonu \times ~
\sum_{j_{23}j_{z23} \alpha}
% ~\sum_{\sigma'}D^{{1 \over 2}} [{\cal R}^\dagger_M(\tilde{\bm \kappa}_{1} )]_{\sigma'\sigma_1}
%D^{{1 \over 2}} [{\cal R}_M({\blf k}'_1)]_{\sigma'_1\sigma'}~
 \delta_{\sigma'_1\sigma_1}~
%\nonu \times ~
{(2\pi)^3~2 k^{'+}_1 }~ \delta^3({\blf k}'_1 -{\blf k}_1^{(a)}) ~ \sqrt{\kappa_1^+ E'_{23}\over k^{'+}_1 E_S} ~
~\sqrt{(2\pi)^3{E'_{23} ~ M'_{23}\over 2 {{M}}'_0(1,2,3)}}~~
\nonu \times ~
\hspace{-0.cm}
\sum_{\sigma_2}\sum_{\sigma_3}
D^{{1 \over 2}} [{\cal R}_M( {\blf k}'_{23})]_{\sigma'_2\sigma_2}~ 
D^{{1 \over 2}} [{\cal R}_M(- {\blf k}'_{23} )]_{\sigma'_3\sigma_3}
~
\langle \sigma_2,\sigma_3; \tau'_2, \tau'_3 ; {\bf k}'_{23}|j_{23},j_{23z};
\epsilon_{23},\alpha;   T_{23}, \tau_{23}
 \rangle  ~ 
 \nonu \times ~ ~
% ~\sum_{\bar{\sigma'}}
%D^{{1 \over 2}*} [{\cal R}^\dagger_M(\tilde{\bm \kappa}_{1} )]_{\bar{\sigma'}\sigma_1}
%D^{{1 \over 2}*} [{\cal R}_M({\blf k}''_1)]_{\sigma''_1\bar{\sigma'}}~
{\delta_{\sigma_1\sigma''_1}}~
%~\nonu \times ~
{(2\pi)^3~2 k^{''+}_1 }~ \delta^3({\blf k}''_1 -{\blf k}_1^{''(a)}) ~ \sqrt{\kappa_1^+ E''_{23}\over k^{''+}_1 E_S} ~
~\sqrt{(2\pi)^3{E''_{23} ~ M''_{23}\over 2 {{M}}''_0(1,2,3)}}~~
\nonu \times ~
\hspace{-0.cm}\sum_{\bar\sigma_2}
\sum_{\bar\sigma_3} 
D^{{1 \over 2}*} [{\cal R}_M( {\blf k}''_{23})]_{\sigma''_2\bar\sigma_2}~ 
D^{{1 \over 2}*} [{\cal R}_M(-{\blf k}''_{23} )]_{\sigma''_3\bar\sigma_3}
~
\langle \bar\sigma_2,\bar\sigma_3; \tau''_2, \tau''_3 ; {\bf k}''_{23}|j_{23},j_{23z};
\epsilon_{23},\alpha;   T_{23}, \tau_{23}
 \rangle ^*  ~ =
 \nonu
 =~2~P'^+~
 (2\pi)^3~\delta ^3({\blf P}'' -{\blf P'}) ~ \delta_{\tau'_1\tau''_1}~
% \sum_{\sigma'}~
% D^{{1 \over 2}} [{\cal R}_M({\blf k}'_1)]_{\sigma'_1\sigma'}~
% D^{{1 \over 2}*} [{\cal R}_M({\blf k}''_1)]_{\sigma''_1{\sigma'}}~
{\delta_{\sigma'_1\sigma''_1}}~
  \int {d{\bf  k}_{1\perp}}\int{{d\xi_1 \over (1-\xi_1)}}
~\nonu \times ~
\hspace{-0.cm}
{(2\pi)^3~ k^{'+}_1 }~ \delta^3({\blf k}'_1 -{\blf k}_1^{(a)}) ~ \sqrt{ E'_{23}\over k^{'+}_1 } ~
~\sqrt{{E'_{23} ~ M'_{23}\over  {{M}}'_0(1,2,3)}}~
\sum_{\sigma_2}\sum_{\sigma_3} ~
D^{{1 \over 2}} [{\cal R}_M{({\blf k}'_{23})}]_{\sigma'_2\sigma_2}~ 
D^{{1 \over 2}} [{\cal R}_M{(-{\blf k}'_{23} )}]_{\sigma'_3\sigma_3}
~
~\nonu \times ~
\hspace{-0.cm}
{(2\pi)^3~ k^{''+}_1 }~ \delta^3({\blf k}''_1 -{\blf {k}}_1^{''(a)}) ~ \sqrt{ E''_{23}\over k^{''+}_1} ~
~\sqrt{{E''_{23} ~ M''_{23}\over  {{M}}''_0(1,2,3)}}~
\sum_{\bar\sigma_2}
\sum_{\bar\sigma_3} ~
D^{{1 \over 2}*} [{\cal R}_M{({\blf k}''_{23})}]_{\sigma''_2\bar\sigma_2}~ 
D^{{1 \over 2}*} [{\cal R}_M{(-{\blf k}''_{23} )}]_{\sigma''_3\bar\sigma_3}
~
\nonu \times ~
\hspace{-0.cm}\sum_{j_{23}j_{z23} \alpha}
\sum_{T_{23} \tau_{23}}\sumint {\lambda(t)~d{  t} } ~\nonu \times ~
\langle \sigma_2,\sigma_3; \tau'_2, \tau'_3 ; {\bf k}'_{23}|j_{23},j_{23z};
\epsilon_{23},\alpha;   T_{23}, \tau_{23}
 \rangle  ~ 
\langle \bar\sigma_2,\bar\sigma_3; \tau''_2, \tau''_3 ; {\bf k}''_{23}|j_{23},j_{23z};
\epsilon_{23},\alpha;   T_{23}, \tau_{23}
 \rangle ^* ~ \quad ,
 \label{overlap1}
 \ee
 where 
 the variable integration change  $d\kappa^+_1 = d\xi_1 ~ E_S/(1-\xi_1)$ was performed (see Eq. (\ref{devkappa})).
  In Eq. (\ref{overlap1}) $~{\bf k''}_{1\perp}^{(a)} = {\tilde{\bm \kappa}}_{1\perp}~$ and 
  $~k''^{+(a)} = \xi_1 ~ \bar M_0(1,2,3)~ $ with 
   \be
 \bar M^2_0(1,2,3) = {m^2 + k^2_{1\perp} \over \xi_1} ~ + ~ {M''^2_{23} + k^2_{1\perp} \over 1 - \xi_1} \quad \quad  .
 \ee
Then, taking into account the completeness for the two-body intrinsic states  
 $\langle \sigma_2,\sigma_3; \tau'_2, \tau'_3; {\bf k}'_{23}|j_{23},j_{23z};
\epsilon_{23},\alpha;   T_{23}, \tau_{23} \rangle $ 
for the (2,3) pair 
(see Eqs. (\ref{nrcompl1}, \ref{nrcompl})), one obtains
\be
A ~ =~2~P'^+~
 (2\pi)^3~\delta ^3({\blf P}'' -{\blf P'}) ~ \delta_{\tau'_1\tau''_1}~
 {\delta_{\sigma'_1\sigma''_1}}~
 %\sum_{\sigma'}~
% D^{{1 \over 2}} [{\cal R}_M({\blf k}'_1)]_{\sigma'_1\sigma'}~
 %D^{{1 \over 2}*} [{\cal R}_M({\blf k}''_1)]_{\sigma''_1{\sigma'}}~ 
 \int {d{\bf  k}_{1\perp}}\int{{d\xi_1 \over (1-\xi_1)}}
~\nonu \times ~
\hspace{-0.cm}
{(2\pi)^3~ k^{'+}_1 }~ \delta^3({\blf k}'_1 -{\blf k}_1^{(a)}) ~ \sqrt{ E'_{23}\over k^{'+}_1 } ~
~\sqrt{{E'_{23} ~ M'_{23}\over  {{M}}'_0(1,2,3)}}~
\sum_{\sigma_2}\sum_{\sigma_3} ~
D^{{1 \over 2}} [{\cal R}_M{({\blf k}'_{23})}]_{\sigma'_2\sigma_2}~ 
D^{{1 \over 2}} [{\cal R}_M{(-{\blf k}'_{23} )}]_{\sigma'_3\sigma_3}
~
~\nonu \times ~
\hspace{-0.cm}
{(2\pi)^3~ k^{''+}_1 }~ \delta^3({\blf k}''_1 -{\blf {k}}_1^{''(a)}) ~ \sqrt{ E''_{23}\over k^{''+}_1} ~
~\sqrt{{E''_{23} ~ M''_{23}\over  {{M}}''_0(1,2,3)}}~
\sum_{\bar\sigma_2}
\sum_{\bar\sigma_3} ~
~
D^{{1 \over 2}*} [{\cal R}_M{({\blf k}''_{23})}]_{\sigma''_2\bar\sigma_2}~ 
D^{{1 \over 2}*} [{\cal R}_M{(-{\blf k}''_{23} )}]_{\sigma''_3\bar\sigma_3}
~
\nonu \times ~
\delta_{\tau'_2,\tau''_2}~\delta_{\tau'_3,\tau''_3}~ \delta_{\sigma_2,\bar \sigma_2}~
\delta_{\sigma_3,\bar \sigma_3}~\delta^3({\bf k}'_{23}-{\bf k}''_{23}) ~ \quad .
\label{overlap2}
\ee

 Therefore, using
 the unitarity of the $D^{1 \over 2}$ matrices and changing the integration variable from $d\xi_1 ~ 1/(1-\xi_1)~$ to $ ~{1 / E'_{23}} ~ dk_1^{+(a)}$ (see Eq. (\ref{deracompl3})),
 one obtains

 \be
\hspace{-.3cm}  A = \delta_{\sigma'_1,\sigma''_1}\delta_{\sigma'_2,\sigma''_2}\delta_{\sigma'_3,\sigma''_3}~
\delta_{\tau'_1,\tau''_1}\delta_{\tau'_2,\tau''_2}\delta_{\tau'_3,\tau''_3}~
2 P^{\prime  +} (2\pi)^9
\delta^3({\blf P}'' -{\blf P}') ~ k'^+_1
%(2\pi)^3
\delta^3({\blf k}''_1 -{\blf k}'_1)~{E^{\prime }_{23}M^{\prime }_{23}\over {{M}}'_0(1,2,3)} 
%(2\pi)^3
\delta^3({\bf k}''_{23} -{\bf k}'_{23})
=\nonu
\hspace{-.2cm} =
\delta_{\sigma'_1,\sigma''_1}\delta_{\sigma'_2,\sigma''_2}\delta_{\sigma'_3,\sigma''_3}~
\delta_{\tau'_1,\tau''_1}\delta_{\tau'_2,\tau''_2}\delta_{\tau'_3,\tau''_3}~
2 P^{\prime  +} (2\pi)^9
\delta^3({\blf P}'' -{\blf P}') ~2 k'^+_1
%(2\pi)^3
\delta^3({\blf k}''_1 -{\blf k}'_1)
~{E'_{23}k_{23}^{\prime+}\over {{M}}'_0(1,2,3)} 
%(2\pi)^3
\delta^3({\blf k}''_{23} -{\blf k}'_{23})=
\nonu
\hspace{-.4cm} =\delta_{\sigma'_1,\sigma''_1}\delta_{\sigma'_2,\sigma''_2}\delta_{\sigma'_3,\sigma''_3}
~\delta_{\tau'_1,\tau''_1}\delta_{\tau'_2,\tau''_2}\delta_{\tau'_3,\tau''_3} ~ 2 P^{\prime  +} (2\pi)^9
\delta^3({\blf P}'' -{\blf P}')   E(k_1')
%(2\pi)^3
\delta^3({\bf k}''_1 -{\bf k}'_1) {E^{\prime }_{23}M^{\prime }_{23}\over {{M}}'_0(1,2,3)} 
%(2\pi)^3
\delta^3({\bf k}''_{23} -{\bf k}'_{23})
\label{compl6}
\ee
The above expressions are  the proper orthogonality relations for the free case, to be related 
to the completeness relations of Eqs. (\ref{compl5}), (\ref{compl3}), and (\ref{compl4}), 
respectively.

\subsection{Product of the non-symmetric basis states and the bound state of the three-particle system}
\label{appc2}
{Let us express} the overlaps between the states of the non-symmetric
 basis (\ref{1e23}) and the bound state of the three-particle system in terms 
 of the canonical wave functions for the two-body and the three-body systems. 
 To this end, {the plane-wave completeness operator (\ref{complpw}) is
 inserted in the intrinsic part of the overlap
 (\ref{overlap}), viz
\be
\hspace{-.4cm} _{LF}\langle   T_{23},\tau_{23};\alpha,\epsilon_{23}; 
j_{23} j_{23z};\tau_1\sigma_1  
\tilde{\bm \kappa}_1 |  j,j_z;
\epsilon^3_{int},\Pi; {1\over 2} T_z \rangle
 =\sum_{\tau_2\tau_3}\sum_{\sigma_2\sigma_3} 
\int {d{\blf  k}'_{23}\over  k^{\prime +}_{23}
(2\pi)^3 }  ~
 \sum_{\sigma'_1}
 \int {{{M}}'_0(1,2,3) ~d{\blf  k}'_1\over 2 k^{\prime +}_1 E'_{23}(2\pi)^3 }
 \nonu 
 \hspace{-.4cm} \times~
 _{LF}\langle   T_{23},\tau_{23};\alpha,\epsilon_{23}; j_{23} j_{23z} ;
 \tau_1\sigma_1  
\tilde{\bm \kappa}_1  | {\blf k}'_{23},\tau_2\tau_3,\sigma_2\sigma_3; 
 {\blf k}'_{1} \sigma'_1\tau_1\rangle_{LF}~
 _{LF}\langle \sigma_2\sigma_3,\tau_2\tau_3, {\blf k}'_{23};\tau_1\sigma'_1  
{\blf k}'_1 |  j,j_z; \epsilon^3_{int},\Pi; {1\over 2} T_z \rangle ~. \nonu 
\ee }
{We can notice that the LF spin states do not change for LF boosts. Therefore the spin states
$|\sigma_2\sigma_3\rangle_{LF}$ in the intrinsic reference frame of the pair (23) or in the intrinsic reference frame of the three-particle system, with momenta related by the LF boost 
$ B_{LF}^{-1}({\blf K}_{23}/M_{23}) $, are equal. Then we can take 
%$ _{LF}\langle   T_{23},\tau_{23};\alpha,\epsilon_{23}; j_{23} j_{23z}  | 
%{\blf k}'_{23},\tau_2\tau_3,\sigma_2\sigma_3\rangle_{LF}$ 
{$ _{LF}\langle   T_{23},\tau_{23};\alpha,\epsilon_{23}; j_{23} j_{23z} ;\tau_1\sigma_1  
\tilde{\bm \kappa}_1   | {\blf k}'_{23},\tau_2\tau_3,\sigma_2\sigma_3;  {\blf k}'_{1} \sigma'_1\tau_1\rangle_{LF}~$ }
as the intrinsic part of the overlap (\ref{lfwf3}) and
$ _{LF}\langle \sigma_2\sigma_3,\tau_2\tau_3, {\blf k}'_{23};\tau_1\sigma'_1  
{\blf k}'_1 |  j,j_z; \epsilon^3_{int},\Pi; {1\over 2} T_z \rangle_{LF} ~ $ as the intrinsic three-body wave function of Eq. (\ref{lfwf4}) and we obtain
}
\be
{\hspace{-.3cm} _{LF}\langle   T_{23},\tau_{23};\alpha,\epsilon_{23}; j_{23} j_{23z};\tau_1\sigma_1  
\tilde{\bm \kappa}_1 |  j,j_z;
\epsilon^3_{int},\Pi; {1\over 2} T_z \rangle }
=\nonu
\hspace{-.3cm} =\sum_{\tau_2\tau_3}\sum_{\sigma_2\sigma_3} ~
\int { d{\blf  k}'_1\over 2 k^{\prime +}_1 (2\pi)^3 }
~\sum_{\sigma'_1}
%\sum_{\sigma'\sigma''}D^{{1 \over 2}} [{\cal R}^\dagger_M({\blf k}'_1 )]_{\sigma'\sigma'_1}
%D^{{1 \over 2}} [{\cal R}_M(\tilde{\bm \kappa}_{1} )]_{\sigma_1\sigma''}~
~
 \int {2{{M}}'_0(1,2,3) ~d{\bf k}'_{23} \over  E'_{23} M'_{23}(2 \pi)^3} ~
{\delta_{\sigma_1\sigma'_1} }~ {(2\pi)^3~2 k^{'+}_1 } ~ \nonu \times ~ 
%({\bf k}_1,\epsilon_{23})~
\hspace{-0,cm}~
 \delta^3({\blf k}'_1 -{\blf k}_1^{(a)})~
\sqrt{\kappa_1^+ E'_{23}\over k^{'+}_1 E_S} ~ 
 \sqrt{{(2 \pi)^3 E'_{23} M'_{23}\over 2 {{M}}'_0(1,2,3)}}
\nonu \times ~
\sum_{\sigma''_2,\sigma''_3}\langle   T_{23},
\tau_{23};\alpha,\epsilon_{23}; j_{23} j_{23z}| {\bf k}'_{23}, \sigma''_2 \sigma''_3,\tau_2\tau_3
\rangle
~  D^{{1 \over 2}} [{\cal R}^\dagger_M{({\blf k}'_{23} )}]_{\sigma''_2\sigma_2}~
D^{{1 \over 2}} [{\cal R}^\dagger_M {(-{\blf k}'_{23} )}]_{\sigma''_3\sigma_3}~\nonu \times ~
~
\sum_{\sigma''_1,\sigma'_2,\sigma'_3}
D^{{1 \over 2}} [{\cal R}_M ({\blf k}'_1 )]_{\sigma'_1\sigma''_1}~
D^{{1 \over 2}} [{\cal R}_M {({\blf k}'_{2} )}]_{\sigma_2\sigma'_2}~ 
D^{{1 \over 2}}
 [{\cal R}_M{({\blf k}'_{3} )}]_{\sigma_3\sigma'_3}
~\nonu \times ~~\sqrt{{ (2\pi)^6~2E({\bf k}'_1) E'_{23} M'_{23} \over  2 {{M}}'_0(1,2,3) }}
~ {\langle \sigma''_1,\sigma'_2,\sigma'_3;\tau_1,\tau_2,\tau_3;{\bf k}'_{23},{\bf k}'_1|j,j_z;
\epsilon^3_{int},\Pi; {1\over 2} T_z \rangle}  \quad .
%\langle{\tilde{\bm \kappa}_1 }| {\blf k}'_{1}\rangle
\label{ovrl2}
\ee
%{\Red where it has been used $\partial k^+_{23}/\partial k_{23 z}=2 /M_{23}$ 
%(cf Eqs. (\ref{derbcompl3}) and (\ref{derbcompl5})).}
{In the previous equation 
%the expressions of Eq. (\ref{lfwf4}) for the three-body LF wave function and of Eq. (\ref{lfwf3}) %have been used and 
the integration variable $k_{23}^{\prime+}$ has been changed in {$k^\prime_{23z}$,  using the equality $\partial k^+_{23}/\partial k_{23 z}=2  k^+_{23}/M_{23}$}
 (see Eqs. (\ref{derbcompl3}) and (\ref{derbcompl5})). Then 
% using the unitarity of the $D^{1/2}$ matrices 
 one obtains }
\be
 _{LF}\langle   T_{23},\tau_{23};\alpha,\epsilon_{23}; j_{23} j_{23z};\tau_1\sigma_1  
\tilde{\bm \kappa}_1 |  j,j_z;
\epsilon^3_{int},\Pi; {1\over 2} T_z \rangle
=\nonu}
{=
\sum_{\tau_2\tau_3}  ~
%\int { d{\blf  k}'_1 }~
{\int d {\bf k}'_{23} ~ 
{~\sum_{\sigma'_1}~
D^{{1 \over 2}} [{\cal R}_M ({\blf k}^{'(a)}_1 )]_{\sigma_1\sigma'_1}~}
\sqrt{{(2\pi)^3 } ~2E({\bf k}^{'(a)}_1)}~
%\delta^3({\blf k}'_1 -{\blf k}_1^{(a)})~
\sqrt{\kappa_1^+ E'_{23}\over k^{'+(a)}_1 E_S}~}
 \nonu
\times{~\sum_{\sigma''_2,\sigma''_3}~\sum_{\sigma'_2,\sigma'_3}\sum_{\sigma_2\sigma_3}~
 D^{{1 \over 2}} [{\cal R}^\dagger_M{({\blf k}'_{23} )}]_{\sigma''_2\sigma_2}~
D^{{1 \over 2}} [{\cal R}^\dagger_M {(-{\blf k}'_{23} )}]_{\sigma''_3\sigma_3}~
D^{{1 \over 2}} [{\cal R}_M {({\blf k}'_{2} )}]_{\sigma_2\sigma'_2}~ 
D^{{1 \over 2}}
 [{\cal R}_M{({\blf k}'_{3} )}]_{\sigma_3\sigma'_3}
 }
~\nonu \times ~
{
\langle   T_{23},
\tau_{23};\alpha,\epsilon_{23}; j_{23} j_{23z}| {\bf k}'_{23}, \sigma''_2 \sigma''_3; \tau_2,\tau_3 \rangle
~
\langle \sigma'_3, \sigma'_2,\sigma'_1; \tau_3,\tau_2,\tau_1; {\bf k}'_{23},{\bf k}^{'(a)}_1|j,j_z;
\epsilon^3_{int},\Pi; {1\over 2} T_z \rangle ~ 
 \quad  
 }{=}
 \nonu
 =
~ \sum_{\tau_2\tau_3}  ~
%\int { d{\blf  k}'_1 }~
 \int d {\bf k}_{23} ~ 
 ~\sum_{\sigma'_1}~
D^{{1 \over 2}} [{\cal R}_M ({\blf k}^{(a)}_1 )]_{\sigma_1\sigma'_1} ~
\sqrt{{(2\pi)^3 } ~2E({\bf k}^{(a)}_1)}~
\sqrt{\kappa_1^+ E_{23}\over k^{+(a)}_1 E_S}\nonu \times 
~
 \sum_{\sigma''_2,\sigma''_3}~\sum_{\sigma'_2,\sigma'_3}
~{\cal D}_{\sigma''_2,\sigma'_2}({\blf k}_{23},{\blf k}_{2})~
{\cal D}_{\sigma''_3,\sigma'_3}(-{\blf k}_{23},{\blf k}_{3})
\nonu \times ~ 
\langle   T_{23},
\tau_{23};\alpha,\epsilon_{23}; j_{23} j_{23z}| {\bf k}_{23}, \sigma''_2 \sigma''_3; \tau_2,\tau_3 \rangle
~
\langle \sigma'_3, \sigma'_2,\sigma'_1; \tau_3,\tau_2,\tau_1; {\bf k}_{23},{\bf k}^{(a)}_1|j,j_z;
\epsilon^3_{int},\Pi; {1\over 2} T_z \rangle ~ 
 \quad  ,
\label{ovrl}
\ee
where
\be
{\cal D}_{\sigma''_i,\sigma'_i}
(\pm{\blf k}_{23},{\blf k}_{i})=
\sum_{\sigma_i}~
 D^{{1 \over 2}} [{\cal R}^\dagger_M{ (\pm{\blf k}_{23} )}]_{\sigma''_i\sigma_i}~
 D^{{1 \over 2}} [{\cal R}_M {({\blf k}_{i} )}]_{\sigma_i\sigma'_i}
 \quad \quad 
\ee
with the $+$  corresponding to $i=2$ and the $-$  corresponding to $i=3$.

Let  us notice that the matrices ${\cal D}_{\sigma''_i,\sigma'_i}(\pm{\blf k}_{23},{\blf k}_{i})$ are unitary, i.e.
\be
\sum_{\sigma_i}{\cal D}^\dagger_{\sigma''_i,\sigma_i}(\pm{\blf k}_{23},{\blf k}_{i}){\cal D}_{\sigma_i,\sigma'_i}(\pm{\blf k}_{23},{\blf k}_{i}) = \delta_{\sigma''_i , \sigma'_i}
\label{uni}
\ee
because of the unitarity of the {${D}^{1/2}$} matrices.

\subsection{Normalization of the overlaps between a state of the cluster $\left\{1,(2 3)\right\}$ and the bound state of the three-particle system}
\label{appc3}
The normalization of the intrinsic LF overlaps
$_{LF}\langle   T_{23}, \tau_{23};\alpha,\epsilon_{23}; j_{23} j_{23z};\tau_1\sigma_1,  {\blf \kappa}_1|  j,j_z; \epsilon^3_{int},\Pi; {1\over 2} T_z \rangle$ 
can be easily recovered using Eq. (\ref{ovrl}), viz
 \be
{N = \sum_{ T_{23}\tau_{23} }
\sumint ~\lambda(t) ~d  t
 \sum_{j_{23}j_{23z} \alpha} 
 \sum_{\sigma_1 \tau_1}  \int {d{\tilde {\bm  \kappa}}_{1}\over 2 \kappa^+_1 (2\pi)^3} 
 \left |_{LF}\langle   T_{23},
\tau_{23};\alpha,\epsilon_{23}; j_{23} j_{23z}; \tau_1\sigma_1,  {\blf \kappa}_1|
  j,j_z; \epsilon^3_{int},\Pi; {1\over 2} T_z \rangle \right|^2=}
\nonu
{=
\sum_{ T_{23}\tau_{23}} 
\sumint ~\lambda(t) ~d{ t}
 \sum_{j_{23}j_{23z} \alpha} 
 \sum_{\sigma_1 \tau_1}  \int {d{\tilde {\bm  \kappa}}_{1}\over 2 \kappa^+_1 (2\pi)^3} 
\left| \sum_{\tau_2\tau_3}  ~
\int d {\bf k}_{23} ~ 
~\sum_{\sigma'_1}~
D^{{1 \over 2}} [{\cal R}_M ({\blf k}^{(a)}_1 )]_{\sigma_1\sigma'_1}~
\sqrt{{(2\pi)^3 } ~2E({\bf k}^{(a)}_1)}~
 \right.}
 \nonu
{\left.
\times ~ {\sqrt{\kappa_1^+ E_{23}\over k^{+(a)}_1 E_S}\sum_{\sigma''_2,\sigma''_3} 
\sum_{\sigma'_2,\sigma'_3}
{\cal D}_{\sigma''_2,\sigma'_2}({\blf k}_{23},{\blf k}_{2})~
{\cal D}_{\sigma''_3,\sigma'_3}(-{\blf k}_{23},{\blf k}_{3})
 }
 \right. }
 \nonu
 {\left.\times ~
\langle   T_{23},
\tau_{23};\alpha,\epsilon_{23}; j_{23} j_{23z}| {\bf k}_{23}, \sigma''_2 \sigma''_3; \tau_2,\tau_3 \rangle
~
\langle \sigma'_3, \sigma'_2,\sigma'_1; \tau_3,\tau_2,\tau_1; {\bf k}_{23},{\bf k}^{(a)}_1|j,j_z;
\epsilon^3_{int},\Pi; {1\over 2} T_z \rangle \right|^2 } ~ =
\nonu
=
\sum_{ T_{23}\tau_{23} }
\sumint {~\lambda(t) ~d{ t}
}
 \sum_{j_{23}j_{23z} \alpha} 
 \sum_{\sigma_1 \tau_1}  
\sum_{\tau_2\tau_3}  ~
%\int { d{\blf  k}'_1 }~
\int d {\bf k}_{23} ~ \int {d{\tilde {\bf k}^{(a)}_1} }
{~\sum_{\sigma'_1}~
D^{{1 \over 2}} [{\cal R}_M ({\blf k}^{(a)}_1 )]_{\sigma_1\sigma'_1}~}
\sqrt{~{E({\bf k}^{(a)}_1) \over E_{23}}}~
 \nonu
\times ~ \sqrt{1\over k^{+(a)}_1}\sum_{\sigma''_2,\sigma''_3} 
\sum_{\sigma'_2,\sigma'_3}
{\cal D}_{\sigma''_2,\sigma'_2}({\blf k}_{23},{\blf k}_{2})~
{\cal D}_{\sigma''_3,\sigma'_3}(-{\blf k}_{23},{\blf k}_{3})
%\sum_{\sigma_2\sigma_3}
% D^{{1 \over 2}} [{\cal R}^\dagger_M{({\blf k}'_{23} )}]_{\sigma''_2\sigma_2}~
%D^{{1 \over 2}} [{\cal R}^\dagger_M {(-{\blf k}'_{23} )}]_{\sigma''_3\sigma_3}~
%D^{{1 \over 2}} [{\cal R}_M {({\blf k}'_{2} )}]_{\sigma_2\sigma'_2}~ 
%D^{{1 \over 2}} [{\cal R}_M{({\blf k}'_{3} )}]_{\sigma_3\sigma'_3}
~\nonu \times ~
\langle   T_{23},
\tau_{23};\alpha,\epsilon_{23}; j_{23} j_{23z}| {\bf k}_{23}, \sigma''_2 \sigma''_3; \tau_2,\tau_3 \rangle
~
\langle \sigma'_3, \sigma'_2,\sigma'_1; \tau_3,\tau_2,\tau_1; {\bf k}_{23},{\bf k}^{(a)}_1|j,j_z;
\epsilon^3_{int},\Pi; {1\over 2} T_z \rangle ~ \nonu \times ~
\sum_{{\bar \tau}_2 {\bar \tau}_3}  ~
\int d {\bf k}''_{23} ~ 
{~\sum_{{\bar \sigma}'_1}~
D^{{1 \over 2}*} [{\cal R}_M ({\blf k}^{''(a)}_1 )]_{\sigma_1{\bar \sigma}'_1}~}
\sqrt{~E({\bf k}^{''(a)}_1)}~
%\delta^3({\blf k}'_1 -{\blf k}_1^{(a)})~
\sqrt{E''_{23}\over k^{''+(a)}_1}~
 \nonu
\times~\sum_{{\bar \sigma}''_2,{\bar \sigma}''_3}~\sum_{{\bar \sigma}'_2,{\bar \sigma}'_3}
{\cal D}^*_{{\bar\sigma}''_2,{\bar\sigma}'_2}({\blf k}''_{23},{\blf k}''_{2})~
{\cal D}^*_{{\bar\sigma}''_3,{\bar\sigma}'_3}(-{\blf k}''_{23},{\blf k}''_{3})
%\sum_{{\bar\sigma}_2{\bar \sigma}_3}~
% D^{{1 \over 2}*} [{\cal R}^\dagger_M{({\blf k}''_{23} )}]_{{\bar \sigma}''_2{\bar \sigma}_2}~
%D^{{1 \over 2}*} [{\cal R}^\dagger_M {(-{\blf k}''_{23} )}]_{{\bar \sigma}''_3{\bar \sigma}_3}~
%D^{{1 \over 2}*} [{\cal R}_M {({\blf k}''_{2} )}]_{{\bar \sigma}_2{\bar \sigma}'_2}~ 
%D^{{1 \over 2}*} [{\cal R}_M{({\blf k}''_{3} )}]_{{\bar \sigma}_3{\bar \sigma}'_3}
~\nonu \times ~
\langle   T_{23},
\tau_{23};\alpha,\epsilon_{23}; j_{23} j_{23z}| {\bf k}''_{23},{\bar  \sigma}''_2 {\bar  \sigma}''_3; 
{\bar \tau}_2,{\bar  \tau}_3 \rangle^*
~
\langle {\bar \sigma}'_3, {\bar \sigma}'_2,{\bar \sigma}'_1; {\bar \tau}_3,{\bar \tau}_2,{\tau}_1; 
{\bf k}''_{23},{\bf k}^{''(a)}_1|j,j_z;
\epsilon^3_{int},\Pi; {1\over 2} T_z \rangle^* ~ \quad.
\label{overlapN2}
\ee
In the last step of Eq.  (\ref{overlapN2}) the change of integration variable 
$d \kappa_1^+ = d k'^{+(a)}_1 {E_S  / E'_{23}}$ (see Eqs. (\ref{deracompl3}) and (\ref{devkappa})) was performed.

Then, using the completeness for the two-body system $(2,3)$ (see Eq. (\ref{nrcompl})) 
%and  the change of integration variable $d \kappa_1^+ = d k'^{+(a)}_1 {E_S  / E'_{23}}$ 
%(see Eqs. (\ref{deracompl3}) and (\ref{devkappa}))
one obtains
{\be
\hspace{-.4cm} N =
 ~  \sum_{\sigma_1 \sigma_2,\sigma_3}  
\sum_{\tau_1\tau_2\tau_3}  ~
%\int { d{\blf  k}'_1 }~
\int d {\bf k}_{23} ~ \int {d{\tilde {\bf k}^{(a)}_1} }~{E({\bf k}^{(a)}_1)\over k^{+(a)}_1}~
\left|{~\sum_{\sigma'_1}~
D^{{1 \over 2}} [{\cal R}_M ({\blf k}^{(a)}_1 )]_{\sigma_1\sigma'_1}~}
%\sqrt{~{E({\bf k}^{'(a)}_1)}}~\sqrt{1\over k^{'+(a)}_1}
\right.  \nonu 
\hspace{-.4cm} \left.
\times ~  
\sum_{\sigma'_2,\sigma'_3}
{\cal D}_{\sigma_2,\sigma'_2}({\blf k}_{23},{\blf k}_{2})~
{\cal D}_{\sigma_3,\sigma'_3}(-{\blf k}_{23},{\blf k}_{3})
~
\langle \sigma'_3, \sigma'_2,\sigma'_1; \tau_3,\tau_2,\tau_1; {\bf k}_{23},{\bf k}^{(a)}_1|j,j_z;
\epsilon^3_{int},\Pi; {1\over 2} T_z \rangle \right|^2 \quad .
\label{overlapN3}
\ee}
{Finally, exploiting  the unitarity of  $D^{1/2}$ and  ${\cal D}$ matrices 
(see Eq. (\ref{uni})), one has }
\be
\hspace{-.4cm} N
= ~  \sum_{\tau_1}  
\sum_{\tau_2\tau_3}  ~
%\int { d{\blf  k}'_1 }~
\int d {\bf k}_{23} ~ \int {d{\tilde {\bf k}^{(a)}_1} }~{E({\bf k}^{(a)}_1)\over k^{+(a)}_1}~
{~\sum_{\sigma'_1}~
}
%\sqrt{~{E({\bf k}^{'(a)}_1)}}~\sqrt{1\over k^{'+(a)}_1}
\sum_{\sigma'_2,\sigma'_3}
\langle \sigma'_3, \sigma'_2,\sigma'_1; \tau_3,\tau_2,\tau_1; {\bf k}_{23},{\bf k}^{(a)}_1|j,j_z;
\epsilon^3_{int},\Pi; {1\over 2} T_z \rangle ~ \nonu \times ~
~
\langle {\sigma}'_3, {\sigma}'_2,{\sigma}'_1; {\tau}_3,{\tau}_2,{\tau}_1; 
{\bf k}_{23},{\bf k}^{(a)}_1|j,j_z;
\epsilon^3_{int},\Pi; {1\over 2} T_z \rangle^* ~ = \nonu
\hspace{-.3cm} {=  \sum_{ \tau_1} 
  \sum_{\sigma'_1}
~ \int d {\bf k}_{23} ~ \int d{\bf {k}}_{1}^{(a)} 
~\sum_{\tau_2\tau_3} \sum_{\sigma'_2,\sigma'_3}
~\left|\langle \sigma'_3, \sigma'_2,\sigma'_1; \tau_3,\tau_2,\tau_1; {\bf k}_{23},{\bf k}_1^{(a)}|j,j_z;
\epsilon^3_{int},\Pi; {1\over 2} T_z \rangle\right|^2
  = 1} \quad  .
\label{overlapN4}
\ee}
where Eqs.  (\ref{deracompl4}) and (\ref{norm3})  were used.

\end{document}